\documentclass[]{aa}  

\usepackage{graphicx}
\usepackage{multirow}
\usepackage{textcomp} 
\usepackage{gensymb}
\usepackage{txfonts}
\usepackage[utf8]{inputenc}
\usepackage[acronym]{glossaries}
\usepackage{natbib}
\usepackage{lineno}
\usepackage{hyperref}
\usepackage{ulem}
\usepackage{svg}
\usepackage{amsmath}
\usepackage{gensymb}

\newcommand{\unsim}{\mathord{\sim}}

\makeatletter
\renewcommand*\aa@pageof{, page \thepage{} of \pageref*{LastPage}}
\makeatother

\begin{document}

   \title{Unveiling the Journey of a Highly Inclined CME} % Identical Interplanetary Coronal Mass Ejection Signatures with Wide Angular Separation

   \subtitle{Insights from the March 13, 2012 Event with\\110$\degree{}$ Longitudinal Separation}

   \author{F. Carcaboso\inst{1, 2, 3}
          \and
          M.~Dumbovi\'c\inst{4}\and
          C.~Kay \inst{2, 3} \and
          D.~Lario\inst{2} \and
          L.~K.~Jian\inst{2} \and
          L.~B.~Wilson~III \inst{2} \and
          R.~Gómez-Herrero\inst{5} \and
          M.~Temmer\inst{6}\and
          S.~G.~Heinemann\inst{7} \and
          T.~Nieves-Chinchilla\inst{2} \and  A.~M.~Veronig\inst{6,8} 
          }

   \institute{
        NASA Postdoctoral Program Fellow, NASA Goddard Space Flight Center, Greenbelt, MD, USA\\
        \email{fernando.carcabosomorales@nasa.gov}
        \and
        Heliophysics Science Division, NASA Goddard Space Flight Center, Greenbelt, MD, USA
        \and
            Physics Department, The Catholic University of America, Washington, DC, USA
       \and 
            Hvar Observatory, Faculty of Geodesy, University of Zagreb, Kaciceva 26, HR-10000, Zagreb, Croatia
        \and
            Universidad de Alcalá, Space Research Group (SRG-UAH), Plaza de San Diego s/n, 28801 Alcalá de Henares, Madrid, Spain 
        \and
            Institute of Physics, University of Graz, Universitätsplatz 5, 8010 Graz, Austria
        \and
            Department of Physics, University of Helsinki, P.O. Box 64, 00014, Helsinki, Finland
        \and
            Kanzelhöhe Observatory for Solar and Environmental Research, University of Graz, Austria
    }
   \date{}

% \linenumbers

% \abstract{}{}{}{}{} 
% 5 {} token are mandatory
 
  \abstract
   % They had a separation of approximately 110 degrees at that moment. As discussed below, Earth crossed the East flank of the ICME and STEREO-A crossed the West one. 
   {A fast ($\unsim2000$ km/s) and wide (>110\degree{}) Coronal Mass Ejection (CME) erupted from the Sun on March 13, 2012. Its interplanetary counterpart was detected in situ two days later by STEREO-A and near-Earth spacecraft, such as ACE, Wind and Cluster. We suggest that at 1 au the CME extended at least 110\degree{} in longitude, with Earth crossing its east flank and STEREO-A crossing its west flank. Despite their separation, measurements from both positions showed very similar in situ CME signatures. The solar source region where the CME erupted was surrounded by three coronal holes (CHs). Their locations with respect to the CME launch site were east (negative polarity), southwest (positive polarity) and west (positive polarity). The solar magnetic field polarity of the area covered by each CH matches that observed at 1 au in situ. Suprathermal electrons at each location showed mixed signatures with only some intervals presenting clear counterstreaming flows as the CME transits both locations. The \textit{strahl} population coming from the shortest magnetic connection of the structure to the Sun showed more intensity.}
   {The aim of this work is to understand the propagation and evolution of the CME and its interaction with the surrounding CHs, to explain the similarities and differences between the observations at each spacecraft, and report what it would be one of the most longitudinal expanded CME structures measured in situ.}
   {Known properties of the large-scale structures from a variety of catalogues and previous studies are used to have a better overview of this particular event. In addition, multipoint observations are used to reconstruct the 3D geometry of the CME and determine the context of the solar and heliospheric conditions before the CME eruption and during its propagation. The graduated cylindrical shell model (GCS) is used to reproduce the orientation, size and speed of the structure with a simple geometry. Also, the Drag-Based Model (DBM) is utilised to understand better the conditions of the interplanetary medium in terms of the drag undergone by the structure while propagating in different directions. Finally, a comparative analysis of the different regions of the structure through the different observatories has been made in order to directly compare the in situ plasma and magnetic field properties at each location.}
   {The study presents important findings regarding the in situ measured CME on March 15, 2012, detected at a longitudinal separation of 110\degree{} in the ecliptic plane despite its initial inclination being around 45\degree{} when erupted (March 13). This suggests that the CME may have deformed and/or rotated, allowing it to be observed near its legs with spacecraft at a separation angle greater than 100\degree{}. The CME structure interacted with high-speed streams generated by the surrounding CHs. The piled-up plasma in the sheath region exhibited an unexpected correlation in magnetic field strength despite the large separation in longitude. In situ observations reveal that at both locations there was a flank encounter, where the spacecraft crossed the first part of the CME, then encountered ambient solar wind, and finally passed near the legs of the structure.} % Also, it provides a possible scenario that explains the differences and similarities between the transit through the different spacecraft.
   {A scenario covering all evidence is proposed for both locations with a general view of the whole structure and solar wind conditions. Also, the study shows the necessity of having multipoint observations of large-scale structures in the heliosphere.
}

   \maketitle
%
%-------------------------------------------------------------------
% \tableofcontents
% \section*{Why this paper (for coauthors) - \textcolor{red}{REMOVE}}

% It is a very wide event (110 degrees in situ) with an identical BSpeed profile and very similar magnetic field strength. Is it the largest ever captured? 

% We can understand the longitudinal differences in the structure, expansion and evolution

% The CME is very tilted, and still, we observe the same structure with that separation. What is causing that? The interaction with the surrounding coronal holes may explain the extreme change in tilt and the drag hampering the propagation towards STA. 

% Despite the proximity between ACE, Wind and Cluster, there is a substantial difference (greater than an hour) in the identification of the regions based on the BSpeed profile. 

% The proposed scenario draws a crossing through the front, then out and then back in through a leg for both locations.

% https://kauai.ccmc.gsfc.nasa.gov/DONKI/view/WSA-ENLIL/1555/1

% \listoftodos

\section{Introduction}\label{sec:introduction}
% CMEs properties and MULTIPOINT observations
Coronal Mass Ejections (CMEs) are large expulsions of plasma originating from the Sun, commonly resulting from processes of magnetic instabilities and reconnection in the solar corona. These structures are constituted by a strongly magnetised plasma which usually remains defined until several astronomical units \citep[e.g.][]{cme_chury} as they propagate quasi-radially outward from the Sun covering a considerable angular extent. Occasionally, the placement of different spacecraft in the heliosphere is appropriate to measure in situ the passage of CMEs at different heliospheric locations. The combination of the corresponding measurements by spacecraft at different heliospheric locations is referred to as multipoint observations. These multipoint observations allow a better understanding of the global and local topology of the large-scale structures, as well as their evolution and  possible interaction with the surrounding medium, as well as with other solar wind (SW) structures \citep[e.g.][]{moestl09,winslow_multipoint}.

The interplanetary counterparts of CMEs are usually divided into distinguishable parts sequentially observed in the following order \citep[e.g.][]{cme_parts}:
\begin{itemize}
\item \textbf{Shock.} Recognised by abrupt plasma parameters increase \citep{shock_summary}. Depending primarily on the CME speed relative to the upstream SW. In the case of subsonic or sub-Alfvenic speeds the shock cannot form.
\item \textbf{Sheath.} Not all CMEs develop sheaths \citep[see e.g.][]{shock_sheath, sheath}. The ambient SW piles up around the CME, compressing it, resulting in an irregular and turbulent plasma and magnetic field. Typically, sheath crossings last $\unsim10$ hours at 1 au \citep{sheath_stats}.
\item \textbf{Magnetic obstacle or ejecta.} A region with an intense magnetic field and commonly few plasma fluctuations. It can be a pre-existing flux rope (FR) or be formed during the eruption of the CME \citep{icme_insitu_signatures, Richardson2010, teresa}. It may also present single or multiple FRs or none. If there is/are FR/s and show/s additional properties like low plasma $\beta$ or bidirectional suprathermal electrons (BDEs), it is often denominated as a magnetic cloud \citep[MC,][]{cross_section_mcs, mc_bde}.
\item \textbf{Post-CME.} After the magnetic obstacle transit, a loosely defined region exists, with mixed characteristics from ambient SW and CME \citep{post_cme}. Signatures like BDEs can still be found \citep{carcaboso2020}. Duration at 1 au can last longer than the actual CME transit \citep{cme_profiles, Temmer_2017, carcaboso2020}.
\end{itemize}

Previous studies such as \citet{vourlidas_catalog} provide some statistics about the properties of the CMEs and estimate that the average angular width of these structures increases with the solar cycle, reaching $\unsim80\degree{}$ as maximum. There are also case-study articles that broadened our knowledge about the complexity of the CMEs, often with the combination of remote sensing observations and in situ measurements. Sometimes, these large-scale structures are captured in situ by multiple spacecraft separated in longitude, which helps to understand the global topology and their inherent dynamics, as well as the interaction with the surrounding SW or other structures. For example, the following studies sorted in longitudinal angular separation analysed different events: \citet{lugaz_55}, 55\degree{}; \citet{74_long_cme}, 74\degree{}; \citet{laura_80}, 80\degree{}; \citet{90_long_cme}, 90\degree{}, and \cite{160_long_cme} 160\degree{}. On those lines, \cite{white_paper_reka} describe the importance of the multipoint analysis to understand these large-scale structures in the SW, while \cite{Scolini_2023} theorises about the amount of spatially-separated spacecraft needed for their complete understanding and characterisation.

Often, the presence of coronal holes (CHs) surrounding the region where CMEs originate, and the SW streams emanating from them (usually, high-speed SW streams) may play a significant role in their propagation. For example, they could act as "magnetic walls" that hamper their propagation \citep{cme_ch_interaction}. The interaction between CMEs and high-speed SW streams can lead to significant changes in CME properties during their journey through the heliosphere, e.g. forcing the internal structure of the CMEs to deform, deflect, kink or rotate \citep{riley_kinematics, he2018stealth, chen2019characteristics, cme_sir_interaction}. Also, high-speed streams could potentially accelerate CMEs, shortening their propagation time from the Sun to Earth \citep{cme_sir_acceleration1, cme_sir_acceleration2, cme_sir_acceleration3}.

% DESCRIBE THE EVENT
A very fast and wide CME erupted from the Sun on March 13, 2012. We discuss the possibility that its interplanetary counterpart was detected in situ two days later by spacecraft located close to Earth (particularly, by the European Space Agency's (ESA's) Cluster \citep{cluster}, and the National Aeronautics and Space Administration's (NASA's) Wind \citep{wind, wind25} and the Advanced Composition Explorer \citep[ACE,][]{ace}) as well as by the Ahead spacecraft of the NASA's Solar Terrestrial Relations Observatory \citep[STEREO-A,][]{stereomission}% , which had a separation of approximately 110 degrees at that moment
. We present a scenario where Earth intercepted the east flank of the CME, whereas STEREO-A, separated by 110\degree{} in longitude with respect to Earth, crossed the west flank. Despite the large longitudinal separation between STEREO-A and Earth, measurements from both positions showed similar in situ CME signatures, especially during the passage of the leading and rear regions of the CME transit at both locations. When the CME was released from the Sun, three CHs were observed surrounding the active region (AR) where the CME originated, resulting in a particular propagation behaviour of this structure towards the different observatories. 

% Their locations with respect to the CME launch site were East (negative polarity), South-West (positive polarity) and West (positive polarity). The polarity on the Sun surface matches with the one observed in situ at 1 au. 

% 

% The CME is observed by all the remote-sensing observatories, and in situ, at least, by STEREO-A and near Earth observatories. These locations were separated $\unsim110$ degrees at the moment of the crossing. 

% Previous studies as \citet{lugaz_55} or \citet{laura_80} broaden our knowledge about the complexity of these structures, revealing the longitudinal differences of the CMEs by the use of multiple spacecraft. Besides, \cite{white_paper_reka} describes the importance of the multipoint analysis of the large scale structures in the SW.

% Big events.... \cite{cme_july2012}

This study aims to explain the propagation and topology of this extremely broad CME released from the Sun on March 13, 2012, and its interaction with the surrounding CHs, using both remote-sensing observations and in situ measurements. Section \ref{sec:observations} shows the evidence from the images provided by the different observatories, and Section \ref{sec:in_situ} shows the local SW properties during the arrival and crossing of the CME. Section \ref{sec:analysis} describes the application of different methodologies supporting the hypothesis that the observed structure resulted from the same solar eruption and were in fact the same CME despite the large longitudinal separation of the locations where the in situ measurements were taken. Section \ref{sec:scenario} outlines a possible scenario consistent with the observations. Finally, Section \ref{sec:discussion} summarises the work and Section \ref{sec:conclusion} discusses the results implied by the scenario that reconstructs the observations.

\section{Remote-sensing Observations}\label{sec:observations}
\begin{figure*}  
   \centering
	\includegraphics[width=\textwidth]{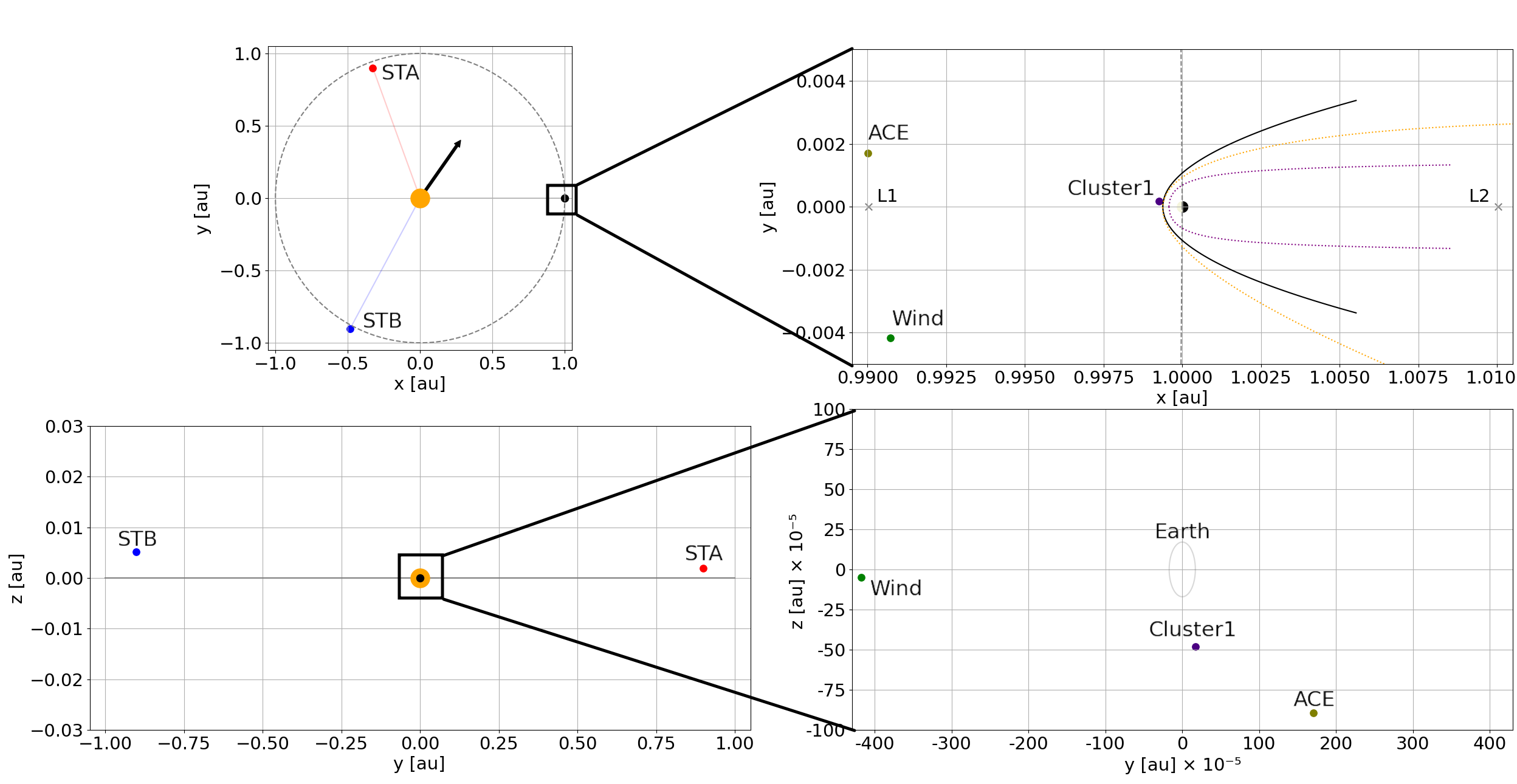}
	\caption{Orbital configuration in HEE coordinates of the analysed spacecraft during the event. Black arrow shows the longitude of the AR at the moment of the eruption. Earth's magnetospheric bow shock (orange) and magnetopause (purple) are calculated as in \cite{bow_shock_model} with a speed of 500 km/s (black line indicates the bow shock without the influence of the SW). Crosses in Earth's perspective indicate the Lagrange points. Earth's radius is scaled by a factor of 4. Sun not to scale. Note the x, y, and z grid sizes are different in some of the panels.
 } 
	\label{fig:orbital_configuration}
\end{figure*}

The locations in the heliosphere of both STEREO spacecraft (STEREO-A and STEREO-B) and near-Earth spacecraft in March 2012 enabled remote-sensing observations that covered the entire Sun in longitude, and thus provided a perfect opportunity to analyse CME structures propagating away from the Sun. %and its identification in the in situ measurements.
The orbital configuration of the spacecraft can be seen in Figure \ref{fig:orbital_configuration}, whereas the coordinates of the spacecraft used in this study, are listed in Table \ref{tab:orbital_configuration}. The instruments used for the remote-sensing observations are listed in Appendix \ref{sec:instrumentation}.

\begin{table}[]
    \centering
    \begin{tabular}{cccc}

s/c & $\phi$ [deg] & $\theta$ [deg]  &   r [au] \\\hline
STEREO-B     & -118.27 $\pm$ 0.12 &  0.29 $\pm$ 0.00 & 1.02  \\
Wind  &  -0.24 $\pm$ 0.00 &  0.00 $\pm$ 0.01 & 0.99  \\
Cluster & 0.01 $\pm$ 0.02 & -0.02 $\pm$ 0.02 &  1.00  \\
ACE     &     0.10 $\pm$ 0.00 & -0.05 $\pm$ 0.00 & 0.99  \\
% MESSENGER*     &  -31.2  &  0.4  & 0.34  \\
% Cluster2 & 0.01 $\pm$ 0.02 & -0.02 $\pm$ 0.02 &  1.00  \\
% Cluster3 & 0.01 $\pm$ 0.02 & -0.02 $\pm$ 0.02 &  1.00  \\
% Cluster4 & 0.01 $\pm$ 0.02 & -0.02 $\pm$ 0.02 &  1.00  \\
STEREO-A     & 110.12 $\pm$ 0.21 &  0.10 $\pm$ 0.00 & 0.96  \\

\end{tabular}
    \caption{Mean orbital position of the spacecraft in Heliocentric Earth-Ecliptic (HEE) coordinate system during the dates 2012-03-14 and 2012-03-20 sorted by longitude, where $\phi$ is longitude, $\theta$ is latitude, and \textit{r} is heliocentric radial distance. Uncertainties correspond to the maximum variation during that period (\textit{r} did not vary significantly).
    %\\*MESSENGER's orbital position is at the day of the hit (2012-03-13).
    }
    \label{tab:orbital_configuration}
\end{table}

In the following subsections, the solar-corona conditions and part of the propagation of the CME captured by the remote-sensing instruments are described. %The interplanetary plasma context where the CME propagated and the measurements acquired during its transit by each spacecraft are described in Section \ref{sec:in_situ}.

% \subsection{Observations of the Source and Solar Conditions}\label{sec:remote_sensing}
% The vantage positions of the spacecraft were used to analyse the active region (AR) where the CME was originated and the CHs in its surroundings. Also, the propagation of the structure thanks to the STEREO-B/HI. 

\subsection{Solar Flare and Coronal Mass Ejection}\label{sec:cme}
% https://www.lmsal.com/isolsearch
The CME under analysis occurred during the rising phase to the solar maximum of solar cycle 24, with multiple events occurring hours before and after \citep[e.g.][]{contradicting_paper, example_march_events}. Noteworthy is the CMEs on March 7, which erupted from the same AR as the CME under analysis and 
caused a widespread solar energetic particle (SEP) event that has been analysed in detail from different approaches \citep[e.g.][]{lario2013, previous_march3, previous_march4, previous_march1, previous_march2}.

Preceding the main CME analysed in this paper, there is a small event seen in STEREO-A/EUVI that happened on March 13 at approximately 15:34 UT from AR11430 located at N22W81 as seen from Earth (catalogued by the National Oceanic and Atmospheric Administration, NOAA, Space Weather Prediction Center, SWPC). This small previous eruption might have triggered the flare and CME eruption under study in this paper due to global instabilities related to large eruptions (e.g., waves, magnetic pressure change, etc.). The full coverage in longitude of the Sun provided by the different spacecraft thanks to their orbital position allows us to confirm that there were no other potential CMEs in the direction of Earth or STEREO-A in a period of 24 hours prior to and after the occurrence of the main CME studied here. 
% \textcolor{red}{mateja.dumbovic: My main concern is that you only make a passing remark on some previous CME, but do not make any attempt to show that there are no CME candidates to arrive at ST-A and Earth besides the analysed CME. We should systematically write all CMEs that occured around our CME, look at their source regions and other basic properties and at least on a descriptive basis make arguments why there are no 'good' candidates other than our CME}

% https://cdaw.gsfc.nasa.gov/movie/make_javamovie.php?img1=sta_e195&img2=sta_co1s&stime=20120313_1200&etime=20120313_1600
% https://cdaw.gsfc.nasa.gov/movie/make_javamovie.php?img1=sdo_a094&img2=sdo_a171&stime=20120313_1501&etime=20120313_1648
% https://cdaw.gsfc.nasa.gov/movie/make_javamovie.php?img1=sta_cor1rd&img2=sta_wav1&stime=20120313_1200&etime=20120313_1600

The main CME under study is related to an M7.9 X-ray solar flare from AR11429 at N18W64 (NOAA/SWPC) %N18° W63° https://solarmonitor.org/index.php?date=20120313&region=11429
with the onset of the soft X-ray emission on March 13 at 17:12 UT. The longitude of this AR is indicated by the black arrow in the top left panel of Figure \ref{fig:orbital_configuration}. The CME was first observed by SOHO/LASCO/C2 at 17:36 UT, and by both STEREO-A and STEREO-B COR1 at $\unsim\text{17:30}$ UT (see Section \ref{sec:cme_reconstruction} for more details). The AR configuration was classified as a complex $\beta\gamma\delta$ \citep[see][and references therein for a detailed explanation]{ar_classification_stats}. Prior to this event, the AR was very dynamic, with multiple CME eruptions \citep[see e.g.][]{liu2013sun, liu2014propagation, Dhakal_2020}\footnote{Images of the event provided by SDO/AIA can be found under the following link: \url{https://sdowww.lmsal.com/sdomedia/ssw/media/ssw/ssw_client/data/ssw_service_120313_105302_91045/www}}.

\citet{erika2018} determined that the CME on March 13 had a left-handed helicity as derived from different multiwavelength proxies (i.e. magnetic tongues X-ray/EUV sigmoids, the skew of coronal arcades, flare ribbons, and filament details. See also \citet{erika2017} and references therein). \cite{Harker_2013} describe in detail the magnetic conditions during the eruption, and the appearance of a magnetic transient during the flare.

\begin{figure}[htbp]  
   \centering
	\includegraphics[width=\columnwidth]{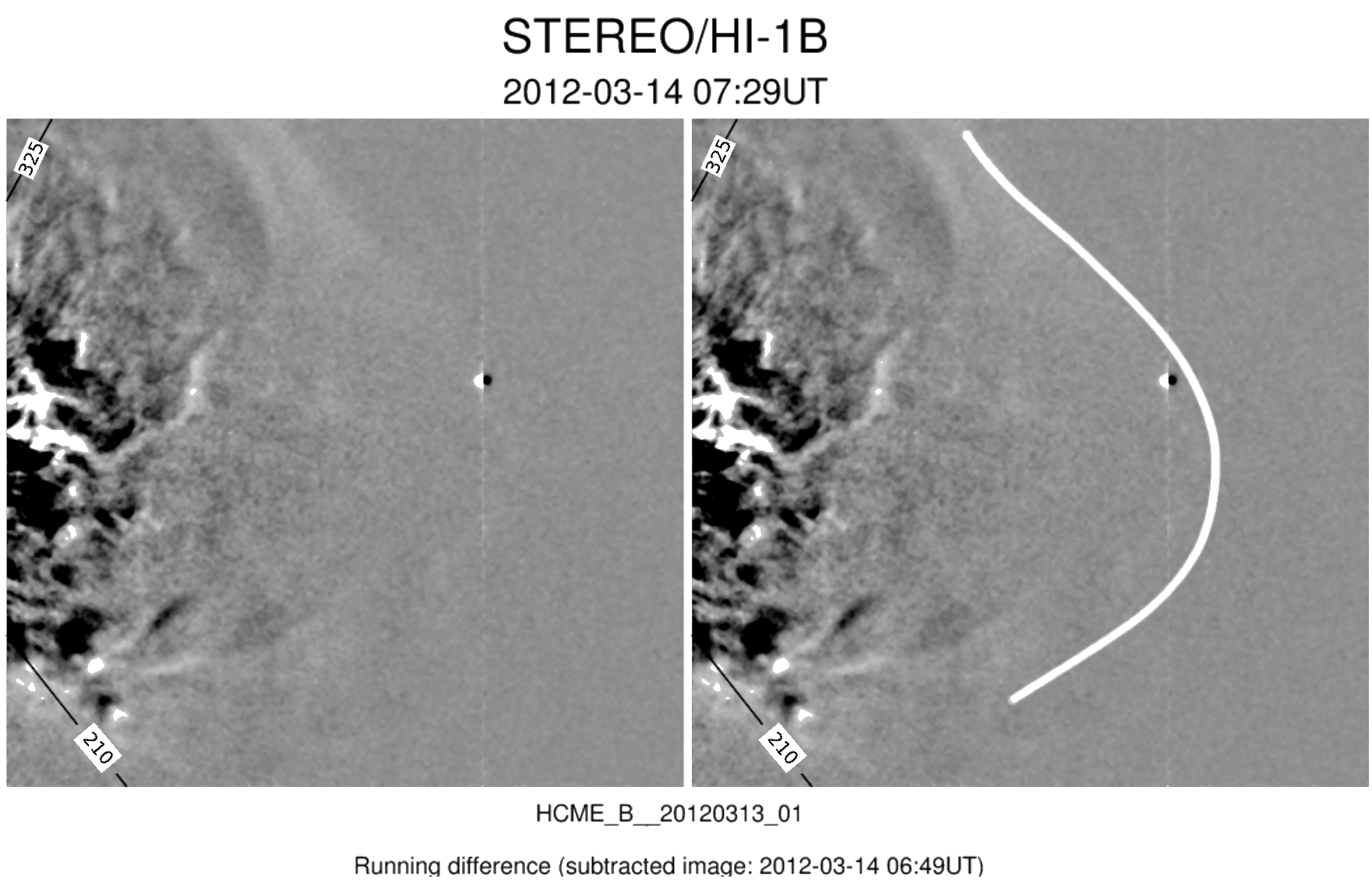}
	\caption{Running difference image from STEREO-B/HI observed on March 14 at 07:29 UT. White line at the right panel indicates approximately the leading edge of the CME. Captured planet (brightest point) corresponds to Mercury. Adapted image from HELCATS.} 
	\label{fig:stereo_b-hi}
\end{figure}

The CME was observed from different white light coronagraphs, appearing as a backside halo CME from STEREO-B's perspective, i.e. moving in the opposite direction of the spacecraft with a main northward propagation. STEREO-A and SOHO observed the CME propagating in the northeast and northwest directions, respectively. STEREO-A/HI and STEREO-B/HI data provided valuable observations of the eastern and western flanks of the wide CME during its heliospheric propagation. The first observation of the structure in STEREO-A/HI and STEREO-B/HI occurred at 18:49 UT and 19:29 UT on March 13, respectively, and was catalogued as \textit{HCME\_B\_\_20120313\_01} and \textit{HCME\_A\_\_20120313\_01} by the Heliospheric Cataloguing, Analysis and Techniques Service (HELCATS)\footnote{\url{https://www.helcats-fp7.eu/catalogues/data/HCME_WP2_V06.txt}}. As seen by STEREO-B/HI, the northern position angle of the CME span was >325\degree{}, while the southern was 210\degree{}. By March 14 at 07:29 UT, the CME's propagating structure showed some distortion on its leading edge, particularly in the northern section, as indicated in the right panel of Figure \ref{fig:stereo_b-hi}.

As a first estimate, extracted from the propagation directions in the coronagraphs and STEREO/HI observations, the CME propagated in the region between Earth and STEREO-A. As reported by HELCATS, the shock speed derived from the Jmap \citep[][]{davies2009Jmaps} is around 1050 $\pm$ 450 km/s. % manuela.temmer: HELCATS derived the shock speed from Jmaps with ca 1050km/s +/- 450km/s. giving a speed with such high precision (643km/s) and then saying that it could also be 1531km/s sounds totally unreliable. the errors are large, we know that. ;-) 
HELCATS also estimates an arrival time at Earth on March 15 at 12:28 UT and with a speed of 850 km/s, which differs by 50 km/s for the speed and $\unsim35$ minutes in advance of the actual arrival time (see ACE, Wind and Cluster observations in Section \ref{sec:in_situ} and Table \ref{tab:timeline} below for more detailed information).

\subsection{Coronal Holes Overview}
\begin{figure*}  
   \centering
	\includegraphics[width=\textwidth]{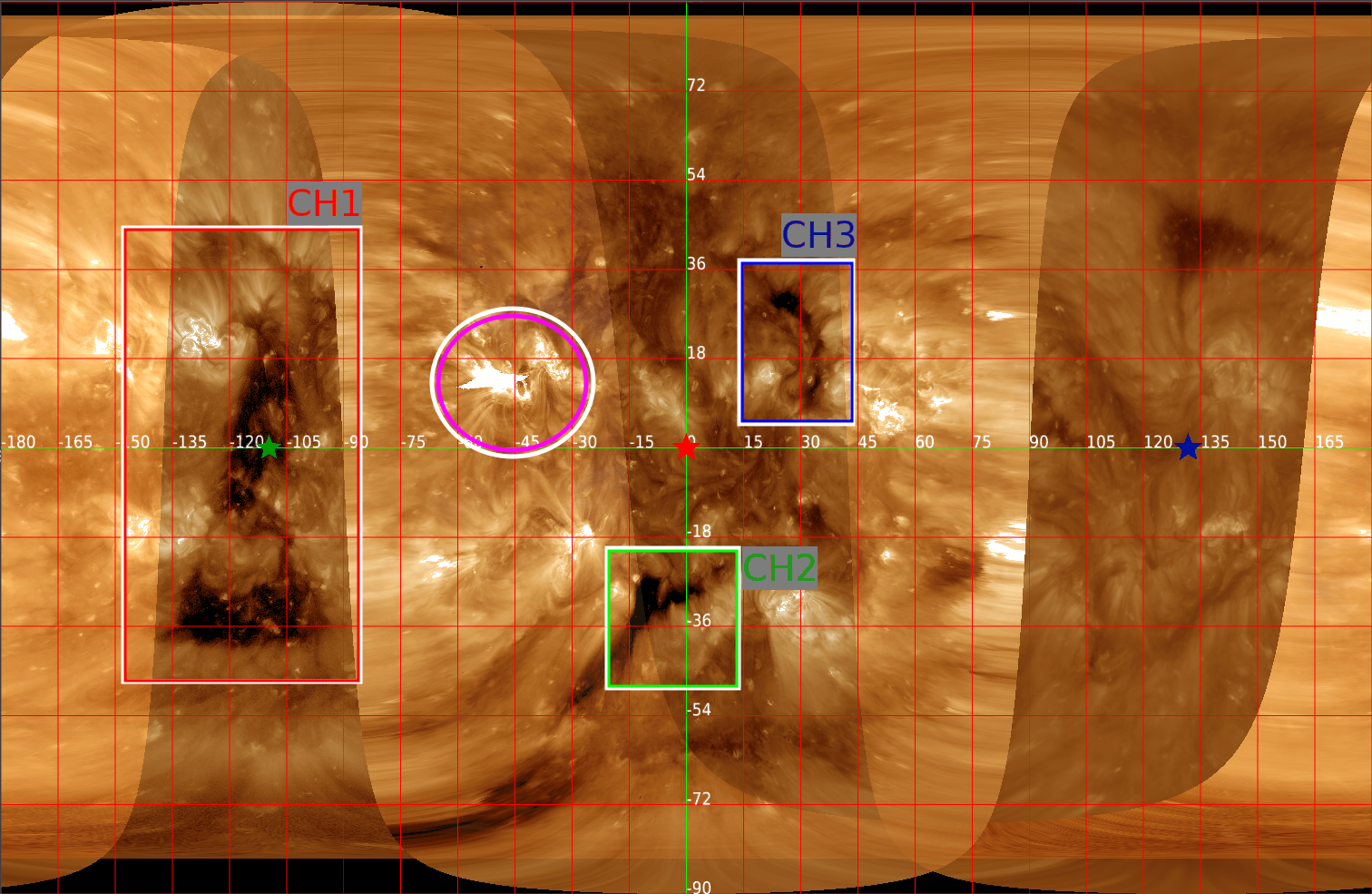}\\
	\caption{Projection of a composed image from SDO/AIA \text{193~\textup{~\AA}}, STEREO-A and STEREO-B EUVI \text{195~\textup{~\AA}} at the moment of the solar flare preceding the CME release. %(2012-03-13T17:35)
	The axes correspond to latitude and longitude as seen from STEREO-A in the ecliptic plane. Pink ellipse shows the AR, while the red, blue, and green boxes indicate the surrounding CHs. Stars represent the projection of each location (green, Earth; red, STEREO-A; blue, STEREO-B). See text for more details.%\\
% 	\textbf{Bottom:} Projection of a composed image from SDO/AIA 193 $\AA$, STEREO-A and STEREO-B EUVI 195 $\AA$ at the moment of the solar flare preceding the CME release.%(2012-03-13T17:35)
	} 
	\label{fig:ch_overview_marked}
\end{figure*}

Three CHs, apart from the northern polar CH, were located surrounding the parent AR. The properties of the CHs were analysed using the program \textit{Collection of Analysis Tools for Coronal Holes} \citep[CATCH,][]{CATCH}. %This tool provides information related to their area, magnetic field intensity, polarity, among others. 
In Figure \ref{fig:ch_overview_marked}, their locations with respect to the parent AR (indicated by a pink circle) can be seen. The East one (CH1, red) had inward (negative) magnetic polarity, while the southwest (CH2, green) and west (CH3, blue) CHs had outward (positive) magnetic polarity. %Their polarity on the Sun surface matches the one observed in situ at 1 au by both Earth and STEREO-A observations. 
At the moment of the eruption, each CH had the following properties: %\textcolor{red}{manuela.temmer: signal location - wave direction, dimming}

\begin{itemize}
    \item \textbf{CH1} was the largest among the three, covering a total area of $\unsim12\times10^{10}~{km}^2$, with its centre of mass at approximately -3\degree{} in latitude and -3\degree{} in longitude (Stonyhurst coordinates. Please note that the reference system is different from the one shown in Figure \ref{fig:ch_overview_marked}). This CH has been studied in more detail, providing information on its evolution and properties by \cite{stephan_ch2, stephan_ch1}.
    \item \textbf{CH2} covered an area of $\unsim4.5\times10^{10}~km^2$, and it is the furthest one with respect to the AR, with its centre of mass located at -36\degree{} (latitude), 99\degree{} (longitude). %As explained in the following sections, it is very unlikely for the CME to have any kind of interaction with this CH, due to its far distance, the presence of the neutral line between them (not shown) and the orientation of the CME.
    \item \textbf{CH3} %is the one most likely interacting with the CME. 
    is the smallest CH, covering an area of $\unsim3\times10^{10}~km^2$ at the moment of the eruption. Its centre of mass was located at 29\degree{} (latitude), 132\degree{} (longitude). %Its closest point in latitude to the AR was found at $\unsim114.92$, which draws an absolute distance of $\unsim60$ degrees.
\end{itemize}

EUV images also show the presence of a diffused region surrounding CH2 and CH3. Due to this reason and the fact that both of them share the same polarity, it could be argued that these two CHs are darker regions within the same large and less-defined CH.

\begin{figure}[htbp]  
   \centering
	\includegraphics[width=1\columnwidth]{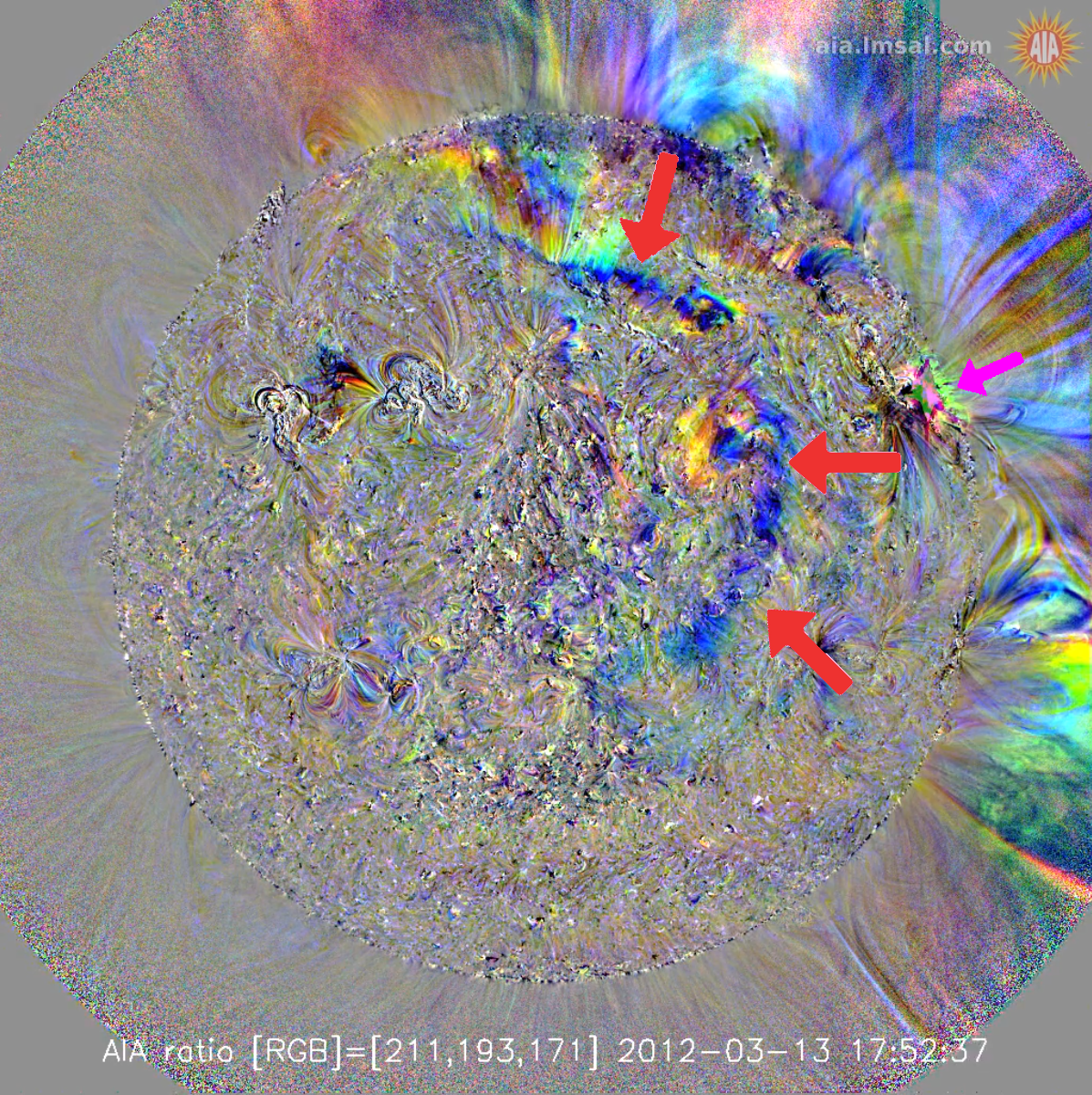}\\
	\caption{Composed running difference image from SDO/AIA \text{211~\textup{~\AA}}, \text{193~\textup{~\AA}}, and \text{171~\textup{~\AA}} observed on March 13 at 17:52:37 UT. Red arrows indicate the appearance of the boundary of CH1 revealed by the interaction with the CME shock wave. The pink arrow points to the AR where the CME originated. %https://iopscience.iop.org/article/10.1088/0004-637X/700/2/L182
    % https://iopscience.iop.org/article/10.3847/1538-4357/ac695c
	} 
	\label{fig:ch_shock_boundary}
\end{figure}

The presence of a low coronal shock wave produced by the CME (see Section \ref{sec:cme}) that clearly interacted with the CHs can be easily observed in the SDO/AIA multi-wavelength composed image\footnote{\url{https://suntoday.lmsal.com/sdomedia/SunInTime/2012/03/13/AIAtriratio-211-193-171-2012-03-13T1200.mov.mp4}}. %http://suntoday.lmsal.com/suntoday/?suntoday_date=2012-03-13
CH1 acted as a barrier hampering the propagation of the shock to further eastern distances, and the boundaries of them are reflected in the video. Figure \ref{fig:ch_shock_boundary} shows a snapshot of that moment at 17:53 UT on March 13, when the EUV wave reflection with CH1 is clearly observed (the red arrows mark the boundaries of CH1, while the pink arrow marks the AR site where the flare occurred). Due to the clear features of this EUV wave and its size, although beyond the scope of this study, a more detailed analysis of the remote-sensing observations of this event in the lower corona is encouraged.

\section{In situ Observations}\label{sec:in_situ}
% \textbf{Decir simplemente lo que se espera, sin decir que es la misma estructura}

Two days after the eruption of the CME, a shock associated with a CME was detected in situ by STEREO-A and by near-Earth spacecraft with a relatively short time difference. At that time, STEREO-A was located at 0.96 au from the Sun and separated 110\degree{} west from Earth (see Figure \ref{fig:orbital_configuration}). In the following subsections, the interplanetary context and the plasma properties during the transit of the relative crossings measured at the different locations are described. The in situ instruments utilised for this purpose are listed in Appendix \ref{sec:instrumentation}. All the represented data have been acquired from the NASA's Coordinated Data Analysis Web (CDAWeb\footnote{\url{https://cdaweb.gsfc.nasa.gov}}), part of the Space Physics Data Facility (SPDF).
% The potential interplanetary counterpart of the CME was measured by spacecraft surrounding Earth and also STEREO-A

\subsection{Near-Earth Observations}\label{sec:in_situ_earth}
\begin{figure*}  
   \centering
	\includegraphics[width=0.49\textwidth]{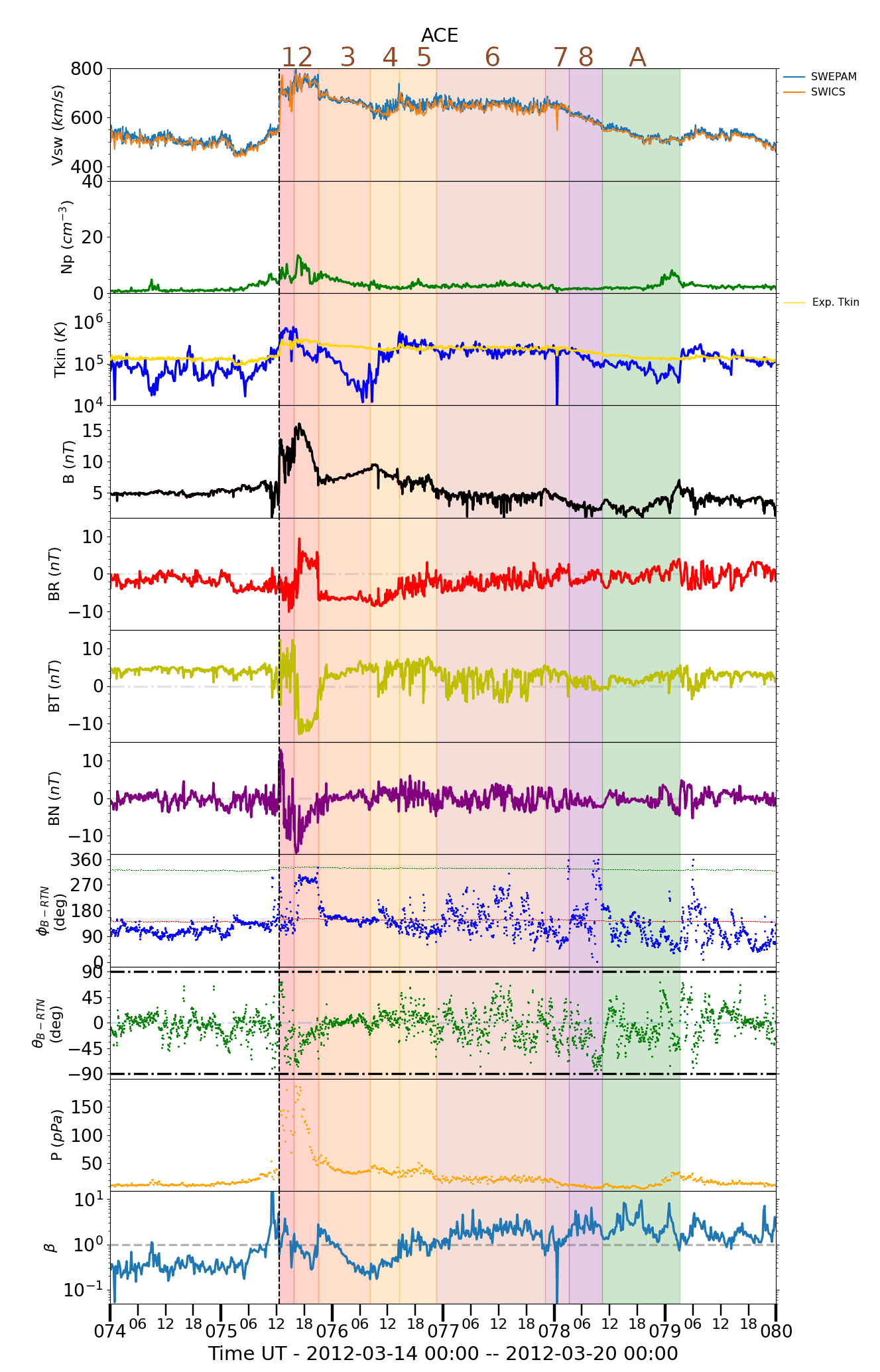}
	\includegraphics[width=0.49\textwidth]{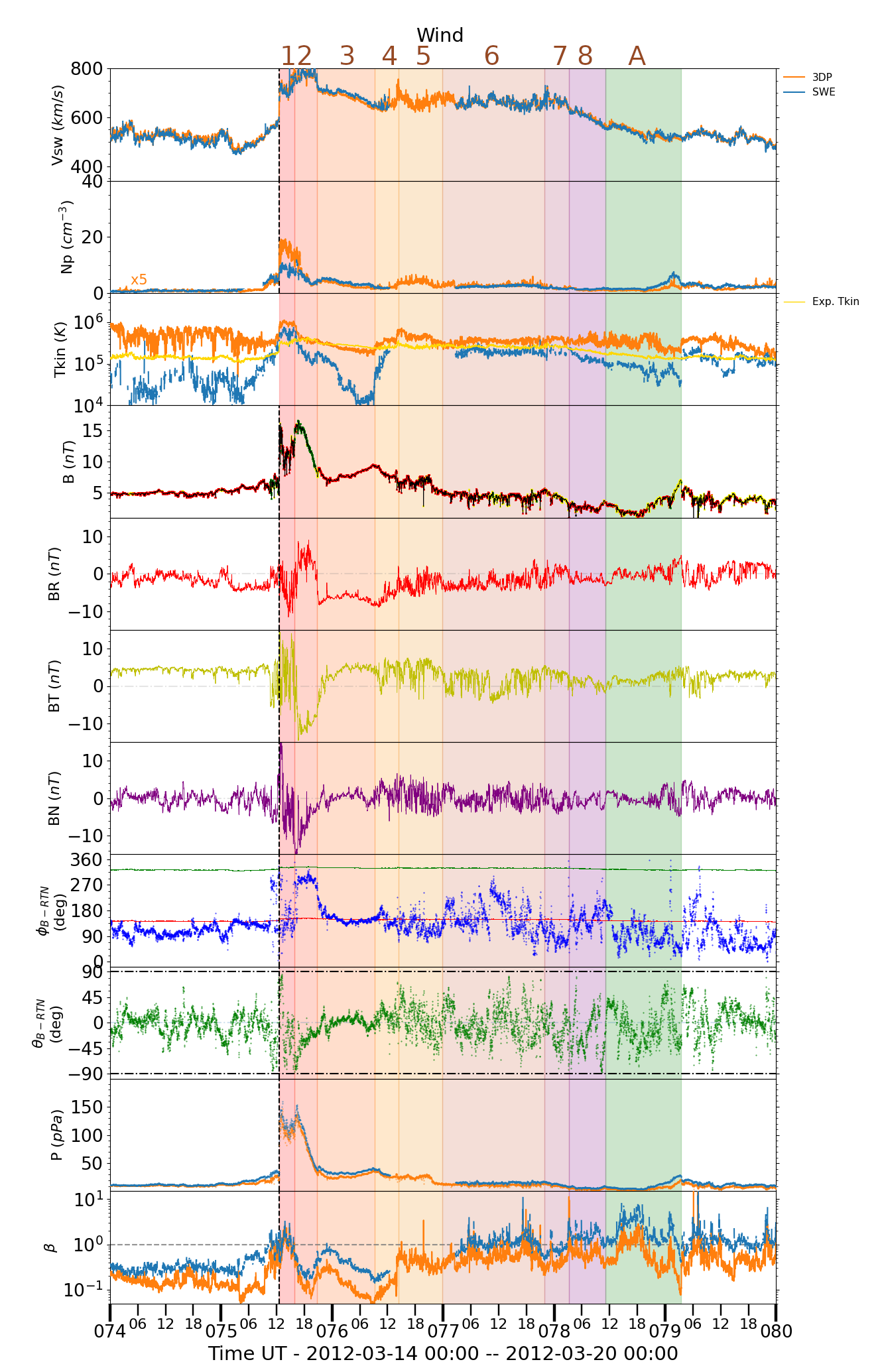}
	\caption{ACE and Wind observations during the period from 2012-03-14 to 2012-03-20. From top to bottom: SW proton speed, proton density, proton temperature, magnetic field magnitude, magnetic field RTN components, magnetic field azimuthal angle in the RTN coordinate system complemented with the two possible nominal Parker spiral angles (red, negative; green, positive) calculated from the proton speed and accompanied by its polarity (red, negative; green, positive; yellow, ambiguous), magnetic field latitudinal angle in the RTN coordinate system, total pressure, and plasma $\beta$.}\label{fig:ace_wind_insitu}
\end{figure*}

Earth's surrounding ambient SW properties were measured by Wind, ACE and the four spacecraft of the Cluster mission. All these spacecraft detected similar time profiles of the magnetic field and plasma parameters (with the exception of Cluster's magnetospheric crossing). %In this case, no shock was catalogued by the Heliospheric Shock Waves Database, of the ICME which is starting on the 16th at XX:XX for ACE, XX:XX for Wind and XX:XX for Cluster.% \noindent \textbf{Wind and ACE}
Figure \ref{fig:ace_wind_insitu} shows the in situ plasma and magnetic field parameters observed by ACE\footnote{During this period, SWEPAM only provided SW speed. SW density and temperature are obtained from SWICS, whereas the total pressure and the plasma $\beta$ parameter are computed using SWICS and ACE/MAG data.} (left) and Wind\footnote{SW plasma parameters are obtained from 3DP (orange) and SWE (blue) data resampled to 1-minute cadence. 3DP density was multiplied by a factor of 5 to be comparable with the other sensors. Due to the low values of density as measured from 3DP, temperature measurements are inaccurate. For this reason, temperature, density and their derivatives from 3DP have not been used for the analysis.} (right) between March 14 (day of year 74) and March 20 (day of year 80), 2012. The panels show from top to bottom: %\textcolor{red}{Lynn -- Be explicit about where data is taken.  That is, are the Wind proton velocity moments coming from the SWE nonlinear fits or the 3DP onboard moments.  If the latter, the data suffer a systematic offset from true with increasing magnitude as time increases from August ~2000 (i.e., the last time those were properly calibrated).  I also like to explicitly state the units in both the text and caption, just for completeness and to help readers with horrible eyesight like myself ;)} 
SW bulk speed, proton density, kinetic proton temperature (blue/orange) together with an empirically predicted temperature (yellow) based on the proton speed as explained in \citet{tkin_speed_orig} and \citet{tkin}, magnetic field strength, its vector components in the spacecraft-centred radial-tangential-normal (RTN) coordinates, the azimuthal angle in the RTN coordinate system (deg) complemented with the two possible nominal Parker spiral angles calculated from the proton speed and accompanied by its polarity (red, negative; green, positive; yellow, ambiguous)\footnote{Nominal Parker spiral angle in the ecliptic plane is calculated as $\phi = arctan(\Omega \cdot r / V_{sw})$, being $\Omega = 2.87\times10^{-6} s^{-1}$. The polarities are determined with a span of $\pm$60\degree{} with respect to the theoretical vectors.}, interplanetary magnetic field latitudinal angle in the RTN coordinate system, total pressure\footnote{Total pressure is calculated as the addition of the plasma thermal pressure ($P_g$) plus the magnetic field pressure ($P_B$). $P_g$ is computed as $P_g = N_p K T_p + N_e K T_e + N_{He} K T_{He}$, where $p\equiv$ protons, $e\equiv$ electrons, \text{$He\equiv$ helium}; $N$ corresponds to density, $K$ to the Boltzmann constant, and $T$ to the temperature. $T_e$ and an alpha/proton ratio are considered constant with a value of 140,000 K and 0.04 respectively \citep{alpha_protons_ratio, electron_temperature, beta_calc}. $P_B$ is computed as ${P}_{B}=B^2 \, /\, 2 \mu_0$, being $B$ the magnetic field strength, and $\mu_0$ the vacuum magnetic permeability.}, and plasma $\beta$.

% manuela.temmer: confusing! maybe explain in the beginning that the subregions are made for a more clear BDE investigation.
The shaded areas indicate different regions based on identifiable bulk speed features, which are associated with different conditions of the SW during the passage of the CME and which may differ from the four distinguishable parts explained in Section \ref{sec:introduction} (see Section \ref{sec:cross_section_comparison} and Table \ref{tab:timeline} for more details). %The exact time of the passage of the different regions are provided below.
The SW conditions preceding the arrival of the CME show no remarkable features, with a SW speed of $\unsim500$ km/s, low density, and constant negative (inward) polarity, almost following the nominal Parker spiral. The vertical black dashed line indicates the passage of the shock associated with the CME. %, and the shaded area is the magnetic obstacle catalogued by Yutian Chi, et al
The shock reaches ACE at 12:31 UT on March 15 (day of year 75), similar to Wind’s time (12:33 UT). The shock parameters are estimated solving numerically the Rankine-Hugoniot relations \citep[e.g.][]{eg_rankine_shock, rankine_koval_szabo} using the \textit{Space Plasma Missions IDL Software Library}\footnote{\url{https://doi.org/10.5281/zenodo.6141586}} \citep{lynntool}. For Wind, the analysis found a quasi-perpendicular shock (the angle between the upstream magnetic field and the normal to the shock, $\theta_B{}_N$, was \text{76.2\degree{} $\pm$ 0.7\degree{}}) with a normal vector <-0.771, -0.538, -0.340> $\pm$ <0.007, 0.005, 0.008> in GSE coordinates (uncertainties are due to statistical errors only; they do not include systematic errors or proper measurement errors), and a fast magnetosonic Mach number of $\unsim3.1$. 
%\textcolor{red}{The components of the vector normal to the shock in XXXXX coordinates are [-0.771, -0.538, -0.340] $\pm$ [+0.007, +0.005, +0.008] for Wind}%\textcolor{red}{ (which according to the interplanetary Heliospheric Shock Database\footnote{\url{http://www.ipshocks.fi/database}} is slightly different: [-0.87, -0.46, 0.19])} and [-0.85, -0.51, 0.11] for ACE in GSE coordinates
% , the angle between the upstream magnetic field and the normal to the shock ($\theta_B{}_N$) of $\unsim76\degree{}$ %\textcolor{red}{($\unsim71\degree{}$)} 
% (i.e. quasi-perpendicular), and a fast magnetosonic Mach number of $\unsim3.1$ %\textcolor{red}{(2.2)} for both of them. 

After the MC, a region with more or less stable SW (with a speed of around 600 km/s) can be found. This region covering the periods \#4, \#5, and \#6 in Figure \ref{fig:ace_wind_insitu} shows higher temperature and lower magnetic field strength (and consequently, higher plasma $\beta$) than the MC. The period is divided into three different parts based on the trend of the SW speed. However, the suprathermal electron pitch-angle distributions (PADs. See Figure \ref{fig:pads} in Section \ref{sec:pad}) show a faint patchy presence of BDEs. This region is more likely to be part of the ambient SW, although there are features that may be related to the MC. According to \cite{contradicting_paper}, these regions correspond to a high-speed stream following the CME which would have ended in region \#3.

Two additional regions (\#7, \#8) correspond to another expanding regime of the SW, with low magnetic field, low proton density, and a variable magnetic field with a predominant negative polarity. The expansion in these two regions is more pronounced than the one in the previous region \#3.

% \textcolor{red}{High iron and oxygen charge states are a good indicator of ICME crossing too \citep{} Bame et al. (1979); Lepri et al. (2001); Lepri and Zurbuchen (2004), Henke et al. (2001); Zurbuchen et al. (2003). During the whole transit of the ICME (from \#1 to \#8), the Fe10+ is very intense, specially during the four first parts (not shown).}

Finally, in contrast to regions \#1 through \#8, region A corresponds to a period for which no correspondence in STEREO-A observations was found (see Section \ref{sec:insitu_sta}). Nevertheless, this region seems to be still part of the expanding structure at Earth and to be interacting with a glancing cross of a SW stream interaction region (SIR) at the end, reduced by the interaction of fast SW generated from likely CH1 and SW left behind the CME, as discussed below. The different parts of this transit are analysed in detail and compared to other spacecraft observations in Section \ref{sec:cross_section_comparison}.

It is also noteworthy the presence of a previous interplanetary CME crossing Earth (not shown) from March 12 at \text{08:28~UT} to March 14 at 02:52 UT, according to the Wind's in situ CME catalogue \citep{wind_icme_catalogue}. This CME may have influenced the upstream conditions of the SW that the CME under study encountered as discussed in Section \ref{sec:analysis}.

% \todo[inline]{more explanation about the PADs}

% The grey dashed line corresponds to shock associated with the ICME, while the green-shaded region shows the corresponding period as in STEREO-A (based on the Vsw and magnetic field profiles) and the grey one corresponds to the one shown in Wind. 

% The magnetic obstacle starts on 17th at 17:35 at Ends on 18th at 19:15

% In this case, no signatures of isotropic suprathermal electrons PAD can be found.

% \begin{figure}  
%    \centering
% 	\includegraphics[width=\columnwidth]{Images/insitu/pads/ACE_PADs.png}
%     \caption{Pitch-angle distribution of the suprathermal electrons of 272 eV measured by ACE/SWEPAM from 2012-03-14 to 2012-03-20. Dashed lines and top colour-bars represent the different defined regions. See text for more details.} \label{fig:ace_pads}
% \end{figure}

The four Cluster spacecraft were out of the magnetosphere during the arrival of the CME (not shown). All four spacecraft detected the shock on March 15 at $\unsim13$:08 UT (in an interval of \text{4.4 seconds} between the first and last observer, 13:07:58.550 UT and 13:08:02.950 UT). Similarly to Wind and ACE, prior to the shock arrival, the SW had negative polarity with a constant magnetic field strength of approximately 6.5 nT. Again, the sheath of the CME could be subdivided into two main parts: a very variable and weak magnetic field (\#1), and a stronger but rapidly decreasing field with with a smoother pointing direction (\#2). The following region (\#3) displays fluctuating magnetic field, but with a clear tendency to increase in strength, until the arrival to the Earth's bow shock at 21:53 UT when the four spacecraft reentered the magnetosphere. The different regions are analysed further in Section \ref{sec:analysis}.
%Figure \ref{fig:cluster_insitu} shows the in situ magnetic field of Cluster1 spacecraft re-sampled to 1 minuted cadence for the period 2012-03-14 to 2012-03-21.
% As Cluster is a magnetospheric mission, the spacecraft are constantly crossing its bow shock and magnetopause. %These periods are grey-shaded in the different panels, which, from top to bottom show the magnetic field strength, RTN components, and the respective angles as explained similarly in Figure \ref{fig:ace_wind_insitu}. Unfortunately, Cluster is unable to provide bulk speed of the SW, so in order to infer the nominal pointing of the Parker spiral within the ecliptic plane, a constant value of 450 km/s has been assumed.

The passage of the CME through the Earth's environment caused a geomagnetic storm with Kp > 5 between 12:00 UT and 21:00 UT on March 15 and reached a minimum equatorial Dst value of -88 nT at 21:00 UT (not shown), while crossing the strongest values of the negative $B_N$ component observed during the passage of region \#2. Then, the Kp index kept a value between 3 and 4 until the first half of March 18, while the Dst was approximately on average \text{-47} nT, and a maximum value of -28 nT for the same period. A Kp = 4 and Dst of $\unsim\text{-40}$~nT were reached between 03:00 and 07:00 UT on March 19, coinciding with the passage of the SIR partially covered in region A (see Figure \ref{fig:ace_wind_insitu}).

% \begin{figure}  
%    \centering
% 	\includegraphics[width=\columnwidth]{Images/insitu/Cluster1-2012-03-14_rd_1.0.png}
% 	% \includegraphics[width=0.24\textwidth]{Images/insitu/Cluster2-2012-03-14_rd_1.0.png}
% 	% \includegraphics[width=0.24\textwidth]{Images/insitu/Cluster3-2012-03-14_rd_1.0.png}
% 	% \includegraphics[width=0.24\textwidth]{Images/insitu/Cluster4-2012-03-14_rd_1.0.png}
% 	\caption{Magnetic field measurements of the four Clusters from 2012-03-14 to 2012-03-21. Figure follows the same format as \ref{fig:ace_wind_insitu}. SW speed considered for the polarity has a constant value of 450 km/s.}
% 	\label{fig:cluster_insitu}
% \end{figure}

% \begin{figure*}  
%    \centering
% 	\includegraphics[width=\textwidth]{Images/insitu/pads/cluster_pads.png}
%     \caption{Pitch-angle distribution of the suprathermal electrons observed by Cluster 1 from 2012-03-14 to 2012-03-21} \label{fig:cluster_pads}
% \end{figure*}

% Figure \ref{fig:cluster_pads} shows the pitch-angle distribution of the suprathermal electrons observed by Cluster1 from the 14th to the 21st for the energy channels: XXXXXXXXXX eV. \todo[inline]{expand}

\subsection{STEREO-A}\label{sec:insitu_sta}

\begin{figure*}  
   \centering
	\includegraphics[width=0.5\textwidth]{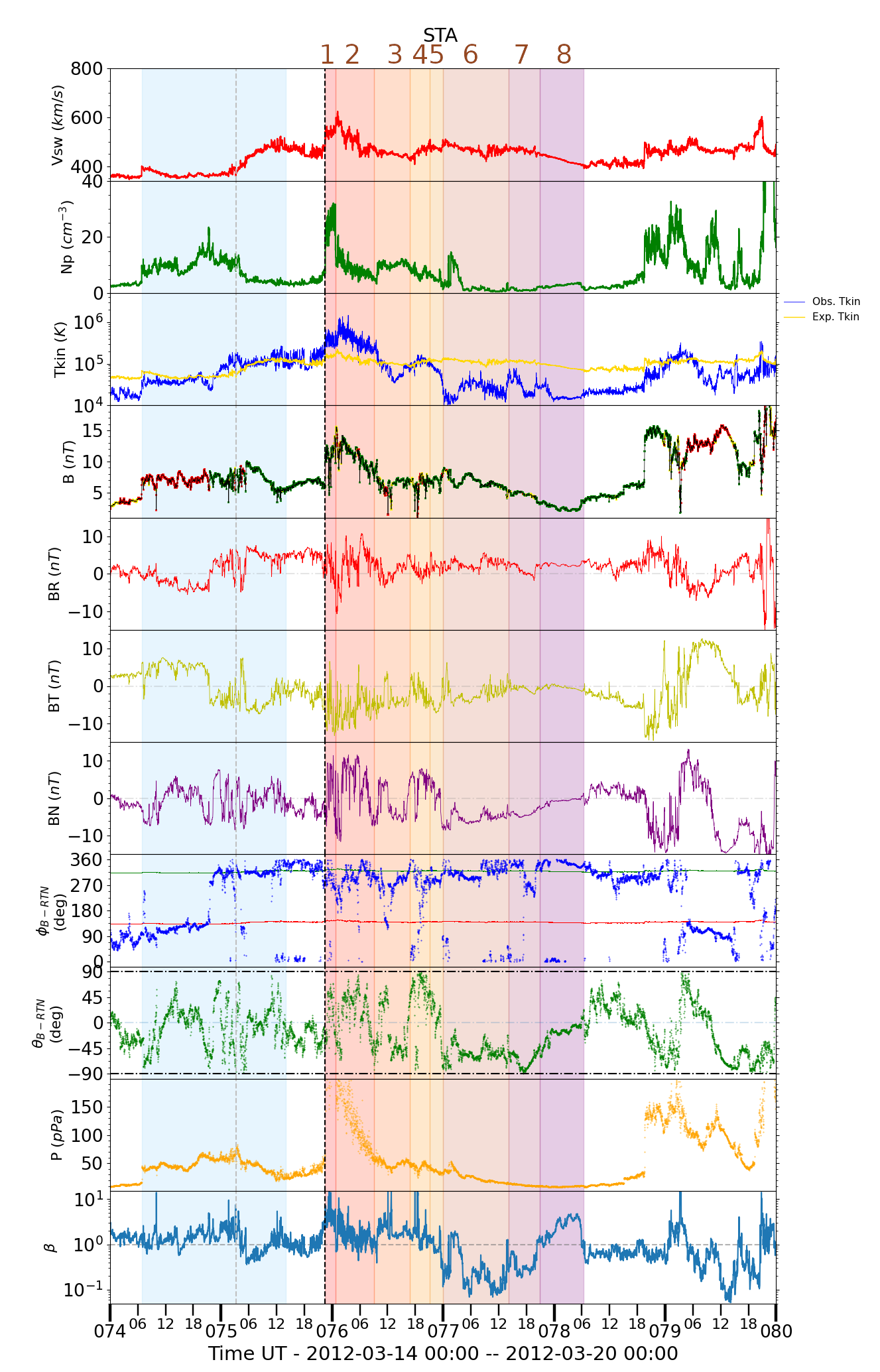}
	\caption{In situ observations from STEREO-A (2012-03-14 to 2012-03-20). Blue-shaded area corresponds to the catalogued SIR (see text for more details). Figure follows the same panel format as in Figure \ref{fig:ace_wind_insitu}.}
	\label{fig:sta_insitu}
\end{figure*}

The crossing of a CME through STEREO-A on the same day (March 15) is also recorded in different in situ CME catalogues (see for example the STEREO/IMPACT catalogue of CMEs\footnote{\url{https://stereo-ssc.nascom.nasa.gov/data/ins_data/impact/level3/ICMEs.pdf}}, listed as the number 8 of year 2012). The criteria for selecting in situ CMEs on that particular catalogue can be found in a series of publications by \cite{Lan_Jian_3, Lan_Jian_2, Lan_Jian_1}. Figure \ref{fig:sta_insitu} displays, using the same format as Figure \ref{fig:ace_wind_insitu}, the STEREO-A in situ measurements for that period. %\textcolor{red}{Fe charge state is high in this CME too. This is surprising for a glancing pass or a pass of one CME leg.} % See https://stereo-ssc.nascom.nasa.gov/data/ins_data/plastic/level2/Iron/plots/Fe_charge_states_daily/PLA_STA_Iron_QStates_20120317.png and https://stereo-ssc.nascom.nasa.gov/data/ins_data/plastic/level2/Iron/plots/Fe_charge_states_daily/PLA_STA_Iron_QStates_20120318.png

%The black-dashed vertical line indicates the shock arrival according to the \textit{Heliospheric Shock Waves Database}\footnote{\url{http://ipshocks.fi}}, which matches also the time of the catalogued CME from the STEREO/MAG catalogue.

The black-dashed vertical line at 22:33 UT on March 15 (day of year 75) indicates the passage of an interplanetary shock. Following the same procedure used for Wind (see Section \ref{sec:in_situ_earth}), the orientation of the shock normal was obtained: <-0.079, -0.997, -0.002> $\pm$ <0.026, 0.002, 0.017> (in GSE coordinates to be directly comparable to the shock normal estimate calculated using Wind data). $\theta_B{}_N$ was very similar to the one measured at Wind (66.5\degree{} $\pm$ 3.7\degree{}, i.e. a quasi-perpendicular shock), and the magnetosonic Mach number was $\unsim2.1$. % -- The shock of the CME, according to the Heliospheric Shock Waves Database, arrived at 22:33%22:32:30
%, with a normal orientation of [0.85, 0.51, -0.12] in HGI coordinates.}
We note that in contrast to near-Earth observations, the shock normal at STEREO-A was mostly in the ecliptic plane with a predominant western direction, whereas at Wind was mostly southwestern directed.

The shock is preceded by a relatively high-speed stream of $\unsim470$ km/s with positive polarity (green horizontal line in the eighth panel of Figure \ref{fig:sta_insitu}), and previous to that, a heliospheric current sheet crossing at 21:30 UT on March 14 (day of the year 74). The increase of SW speed observed early on March 15 preceded by a period of elevated SW density and magnetic field magnitude late on March 14 can be interpreted as the arrival of a SIR, which presumably originated from CH2 or CH3, as their longitude and polarity agree with the observed features. According to the Predictive Science Inc. (PSI) magnetic back-mapping results\footnote{\url{https://www.predsci.com/hmi/spacecraft_mapping.php}} (not shown), STEREO-A is more likely connected to the northern hemisphere on March 15, where CH3 is located. The period with elevated magnetic field and density is also reported in the STEREO/IMPACT catalogue of SIRs\footnote{\url{https://stereodata.nascom.nasa.gov/pub/ins_data/impact/level3/STEREO_Level3_SIR.pdf} See \citet{sir3, sir1} for the selection criteria.} (shaded in blue). The time of the stream interface passage according to the catalogue is marked by the vertical dashed grey line in Figure \ref{fig:sta_insitu} within that area, and corresponds to the highest pressure during the passage.  %According to the catalogue, the boundaries of the SIR are between XXXXX and XXXX, while the stream interface is crossed at XXXXXX. % Nevertheless, from our perspective, there is no clear signatures in the magnetic field nor plasma parameters to define at those boundaries.

Similarly to the observations at Earth, the sheath (region \#1) has a variable but intense magnetic field strength, and high proton density and temperature, and it is followed by another region with an even more intense and smoother magnetic field magnitude, and less dense too (region \#2). Both of them show elevated temperatures and have a high plasma $\beta$. Right after them, an expanding region (\#3) with low temperatures and constant magnetic field strength (around 7 nT) can be found. Later on, a period with more or less constant SW speed followed the previous regions, with even lower proton density on average (corresponding to regions \#4, \#5, \#6). These regions are even colder and less dense than the previous ones, their magnetic field is decreasing in strength and is pointing with an inclination of $\unsim45\degree{}$ with respect to the ecliptic with a main positive (outward) polarity. Consequently, the plasma $\beta$ in region \#6 is lower than the surroundings.

The last part (regions \#7, \#8) is another expanding regime of the SW, with a very weak magnetic field and similar plasma properties as the previous regions, which draws a higher plasma $\beta$. The magnetic field is weak but smooth, going from a south-pointing direction to the nominal behaviour of the Parker spiral within the ecliptic plane.

A few hours after this last region (on March 18 at \text{$\unsim$19:30~UT}), another interplanetary shock produced by a different CME arrives at the spacecraft (not indicated in Figure \ref{fig:sta_insitu}).

% El MO no corresponde con el que pensamos nosotros: Nevertheless, the MO obstacle of the structure, as established by the catalogued, is shaded in green between 2012-03-15TXX:XX and 2012-03-15TXX:XX. % vaya mojon
% Although the catalogue tries to have an accurate reconstruction of the structures, a deeper analysis is always encouraged. 
% Y proponemos otro que se puede ver en la seccion donde se comparan:
% In this work, a different perspective of the structure is suggested as seen in section \ref{sec:cross_section_comparison}.

% (XX) panel shows the magnetic field
% (XX) panel shows the plasma bulk speed
% (XX) panel shows the components
% (XX) panel shows the magnetic field
% (XX) panel shows the magnetic field
% (XX) panel shows the magnetic field

% \url{https://ccmc.gsfc.nasa.gov/ungrouped/search_main.php}

A more detailed study of the remote-sensing observations and in situ measurements is performed in the following sections aiming to explain the most probable scenario providing evidence that the structure intercepted at both locations with $\unsim110\degree{}$ of longitudinal separations corresponds to the same CME that erupted from the Sun on March 13.

\subsection{Other locations}
%manuela.temmer: I would erase that and say that additionally MESSENGER data are available. Make it a strength that you add one more in situ and not a weakness that you could not add 5 more ;-)
% mateja.dumbovic: Does this info in any way help our analysis, if not I would remove it (especially as we kind of leave it on a speculative basis?) but if we decide to stic´k with this messenger should be added to the spacecraft constellation plot
We have considered other missions that were also located in the trajectory of the CME or at least in its vicinity and may have been potentially impacted. Nevertheless, they are not shown in the present article due to the complexity of the analysis of their data in addition to evolutionary processes which are out of the scope of the paper. 

As a note, one of them was the NASA's Mercury Surface, Space Environment, Geochemistry and Ranging (MESSENGER) mission \citep{messenger} which was orbiting about Mercury. MESSENGER magnetometer detected the passage of an interplanetary shock at 07:08 UT on March 14 with clear signatures of the passage of a CME at the moment that MESSENGER was in the SW (i.e. outside of Mercury's magnetosphere). %  and clear signatures of the transit of an ICME, when excluding those periods of crossing the planetary magnetosphere.
At that moment, the spacecraft was located at -31.2\degree{} in longitude, 0.4\degree{} in latitude and 0.34 au (HEE). Owing to the absence of previous solar events, the observing time ($\unsim8.5$ hours after the eruption) and the MESSENGER location, we suggest that the observed structure corresponds to the same CME under study. Nevertheless, the in situ CME has not been catalogued in previous studies \citep[e.g.][]{messenger_catalogue}. Note that the longitudinal separation between the site of the parent eruption and the MESSENGER spacecraft was $\unsim85\degree{}$, which is not inconsistent with the typical width of CME-driven shocks in the interplanetary space \citep[e.g.][]{shock_extent_multipoint}.

\section{Data Analysis}\label{sec:analysis}
This section focuses on the morphology and dynamics of the CME from its solar origin to the in situ interplanetary measurements, assuming that the in situ measured structure at Earth and STEREO-A, with a relative separation larger than 100\degree{}, corresponds to the same CME. We also show that the structure deflected from an initial considerable out-of-the-ecliptic inclination, to be observed in the ecliptic plane with that separation. %; 3) the CME experienced interaction with a high-speed stream in the interplanetary medium, hindering its propagation towards STEREO-A.

\subsection{Reconstruction of the CME near the Sun} \label{sec:cme_reconstruction}

% mateja.dumbovic: In this section you should also compare remote and in situ properties of the FR, e.g. the orientation from GCS, ST-A and ACE and Wind. Also you should comment the longitudinal spread (width) of the CME based on the remote observation and how does it compare to spacecraft separation? Here you should address the problem of peculiarity of this CME which appears relatively normal from remote sensing but is then detected in situ at longitudinally greatly separated spacecraft.

\begin{figure}[htbp]  
   \centering
	\includegraphics[width=0.32\columnwidth]{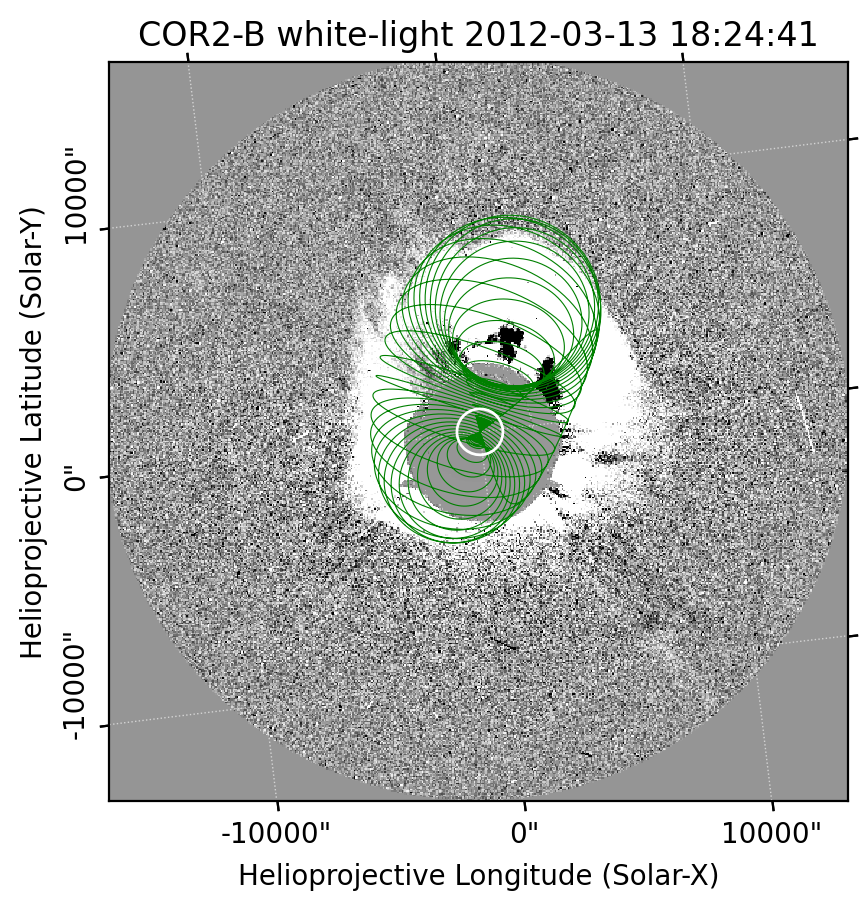}
 	\includegraphics[width=0.33\columnwidth]{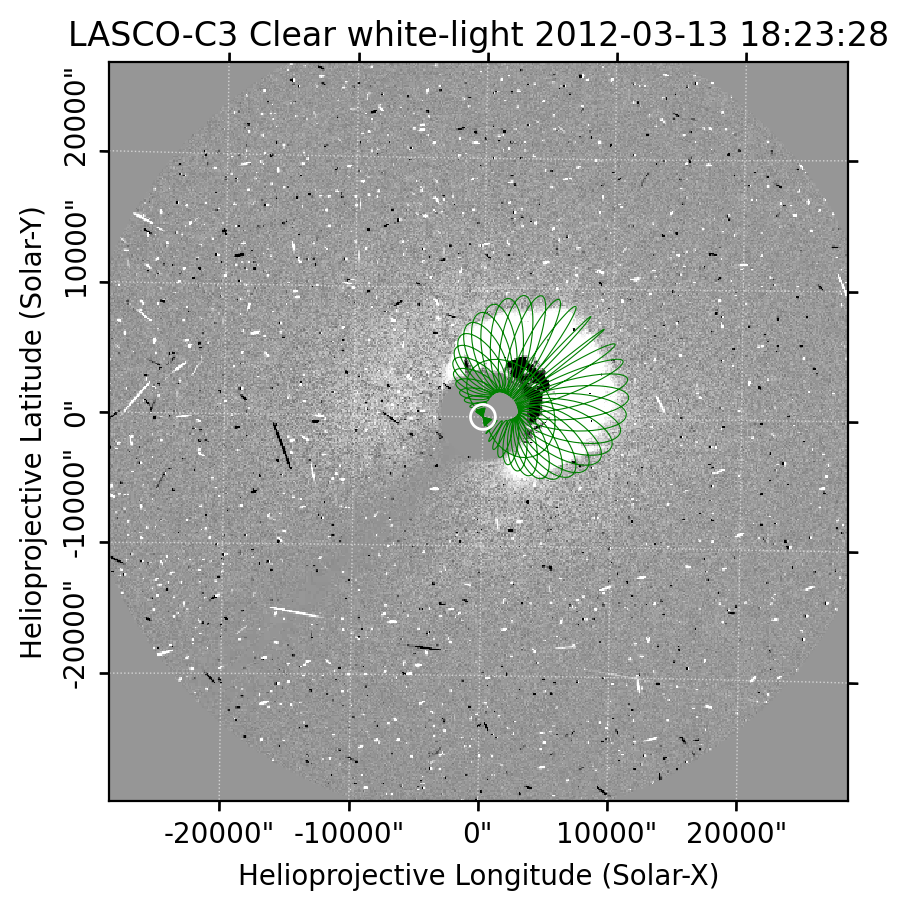}
 	\includegraphics[width=0.32\columnwidth]{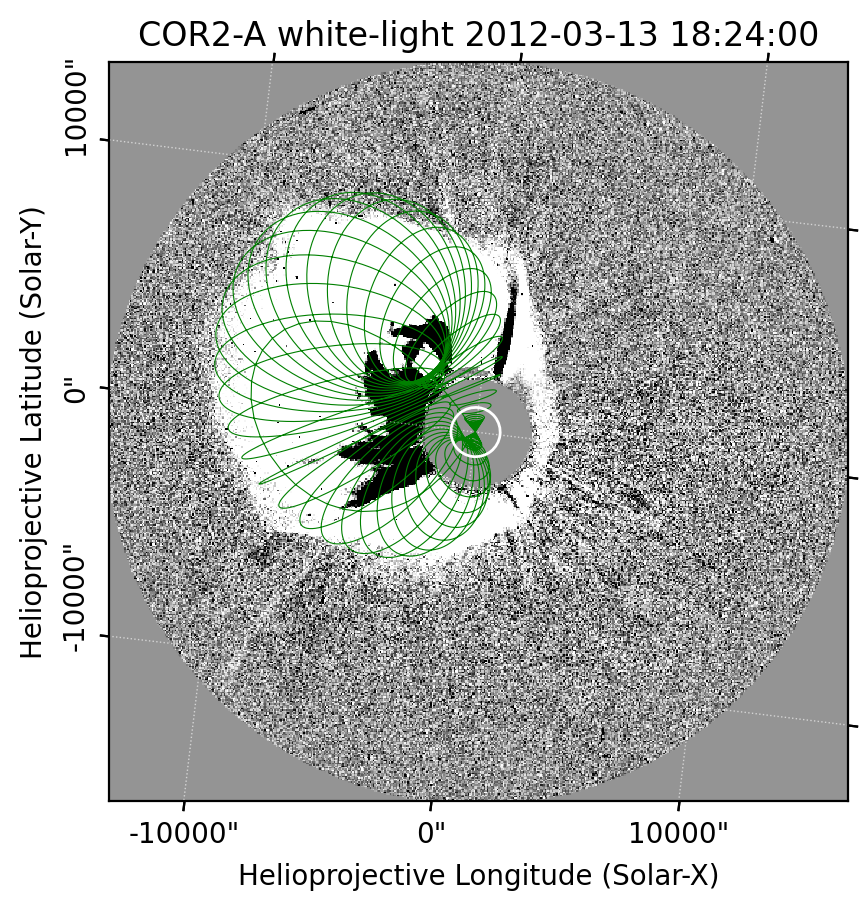} 	
  \includegraphics[width=0.5\columnwidth]{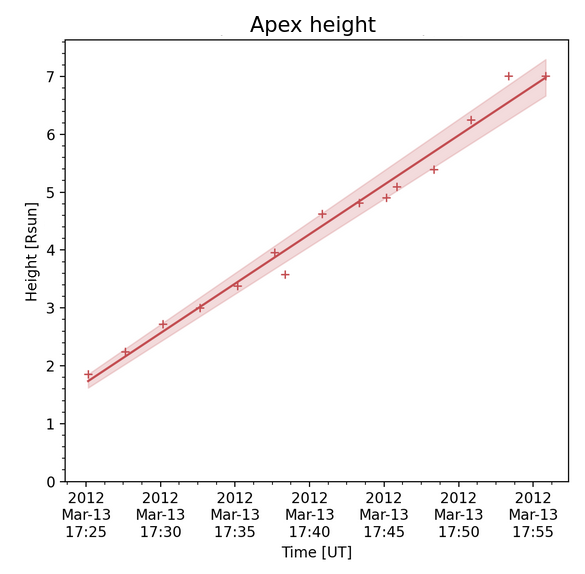}
	\caption{\textbf{Top:} Running difference images provided by STEREO-B/COR2 (left), SOHO/LASCO/C2 and STEREO-A/COR2 (right) on 2012-03-13 at $\unsim18:24$ UT. GCS reconstruction is shown in green. See text for more details. \textbf{Bottom:} Height of the apex as a function of time along with linear fit to the data. %\textcolor{red}{As the fit is a linear fit , i.e. const speed, the right panel can be removed since their is no speed evolution but only a single value.Instead, make the left panel bigger, and annotate the derived mean speed in this panel.}
 } \label{fig:gcs}
\end{figure}

For the reconstruction of the size, orientation and kinematics of the structure, the graduated cylindrical shell (GCS) geometrical model has been used \citep{GCS} utilising the Python-based PyThea tool \citep{pythea}. This model assumes a structure similar to a croissant in shape composed of two cones and part of a torus, where different parameters can be adjusted to reconstruct the coronagraph observations, and thus reproduce the leading edge of the structure and obtain, among others, its 3D shape and orientation. A snapshot of the fits applied simultaneously to STEREO-A, STEREO-B and SOHO coronagraph observations is shown in the top row of Figure \ref{fig:gcs}. The bottom panel of Figure \ref{fig:gcs} shows the derived height of the leading edge of the GCS model (i.e. apex) as a function of time, providing a speed of 1982 $\pm$ 75 km/s when fitted to a first-order function, similar to the one listed in the SOHO/LASCO catalogue (1884--2054 km/s)\footnote{\url{https://cdaw.gsfc.nasa.gov/CME_list/UNIVERSAL/2012_03/univ2012_03.html} . The two values 1884 km/s and 2054 km/s correspond to a linear and second-order fit to the leading edge of the CME (position angle 286\degree{}) versus time as seen in plane-of-sky LASCO/C2 and LASCO/C3 observations.}. The GCS results show an average orientation of $\unsim18\degree{}$ in latitude and $\unsim55\degree{}$ in longitude in Stonyhurst coordinates, a tilt angle of approximately -45\degree{} measured counterclockwise positive from the Earth-Sun direction. A similar orientation was obtained in the study \citet{erika2018}. Our GCS reconstruction is consistent in longitude and latitude to the AR location on the Sun (N18W64). Also, despite the observed interactions in the lower corona with the surrounding CHs (see Section \ref{sec:cme}) and the high-speed streams coming from them, the direction and orientation of the CME did not considerably change in the coronagraphs' field of view.

% These inputs were used for the ENLIL simulation, and can be found under the link \url{https://kauai.ccmc.gsfc.nasa.gov/DONKI/view/CMEAnalysis/1556/1}.

% \url{https://kauai.ccmc.gsfc.nasa.gov/DONKI/view/WSA-ENLIL/1555/1}

 % mateja.dumbovic: GCS plots are too small, I would stretch the figure across the whole page. Also the numbers on axis are way too small I would either increase them or remove them

\subsection{CME Propagation}

% \textcolor{red}{mateja.dumbovic: what CME input was used?? From GCS? what is the width? are both flank hits? More info on how you run DBM is needed here}

% \textcolor{red}{manuela.temmer: can you infer from the orientation of CME and CH whether reconnection would have been likely? }

Taking into account the arrival time of the shock at both locations (see Section \ref{sec:in_situ}) and that the eruption appeared on the coronagraphs at \text{$\unsim\text{17:40}$~UT} on March 13 as indicated by the onset of the soft X-rays, the transit time of the CME to reach Earth is $\unsim42.96$ hours, while for STEREO-A it takes $\unsim52.88$ hours (both times have been computed considering that the eruption started at the onset time of the soft X-ray emission of the associated flare). Also, according to the GCS reconstruction (see Section \ref{sec:cme_reconstruction}), the CME centre (i.e. apex) was propagating \text{$\unsim55\degree{}$} apart from both Earth and STEREO-A. In order to have an estimation of the drag undergone by the CME while propagating in the interplanetary medium, the 1D Drag-Based Model (DBM)\footnote{Available at \url{https://oh.geof.unizg.hr}} has been used \citep{dbm, dbm_tools}, which provides a drag parameter ($\gamma$) based on the propagation time. By using the in situ measured SW speed, and assuming that the speed of the structure was the same in both directions and similar to the apex ($\unsim2000$ km/s) %You: where does this speed come from? in the section above you state about 2000 km/s. Also, the speed of the CME segments toward the s/c may be different in the different insitu s/c directions/. Also give the diferent solarwind speeds that you have used. This all needs clear elaboration here, how the calculation was done and which numbers for the paramters were used and how derived.
, $\gamma$ obtained for Earth is $0.3753\times10^7 km^{-1}$, while for STEREO-A is $0.5464\times10^7 km^{-1}$, suggesting that the structure underwent more drag ($\unsim46 \%$ more) during its propagation towards STEREO-A than along the Sun-Earth line. 

There are two distinct processes that may explain the difference in drag although the CME propagated with similar longitudinal separation with respect to both spacecraft. These are the swept ambient SW in the Sun-Earth line due to a previous CME (see Section \ref{sec:in_situ}) and the presence of the SIRs produced by CH2 and CH3. The less-dense conditions left behind the previous CME allow a faster propagation of the CME under study towards Earth \citep{liu2014, prop_july2012}, while the presence of SIRs hampers the propagation in STEREO-A direction \citep{cme_ch_interaction}. This can also be appreciated in the higher proton temperature at STEREO-A than at Earth during the first part of the structure, and the difference in density in the upstream SW ($\unsim5~cm^{-3}$ for Earth, $\unsim8~cm^{-3}$ for STEREO-A), but especially during the transit of the sheath ($\unsim9.5~cm^{-3}$ for Earth, $\unsim27~cm^{-3}$ for STEREO-A. See Figure \ref{fig:superposed_epoch}), which is more representative of the conditions that the CME experienced during its evolution through the interplanetary medium.

\subsection{Longitudinal SW Comparison}\label{sec:cross_section_comparison}
There is clear evidence suggesting that the plasma measurements at both locations (i.e. at STEREO-A and near-Earth) correspond to different parts of the same structure. In order to prove this, in this Section we perform a more detailed comparison of the plasma properties measured at the subdivided regions of the passage of the CME at both locations.

\begin{figure}[htbp]  
   \centering
	\includegraphics[width=\columnwidth]{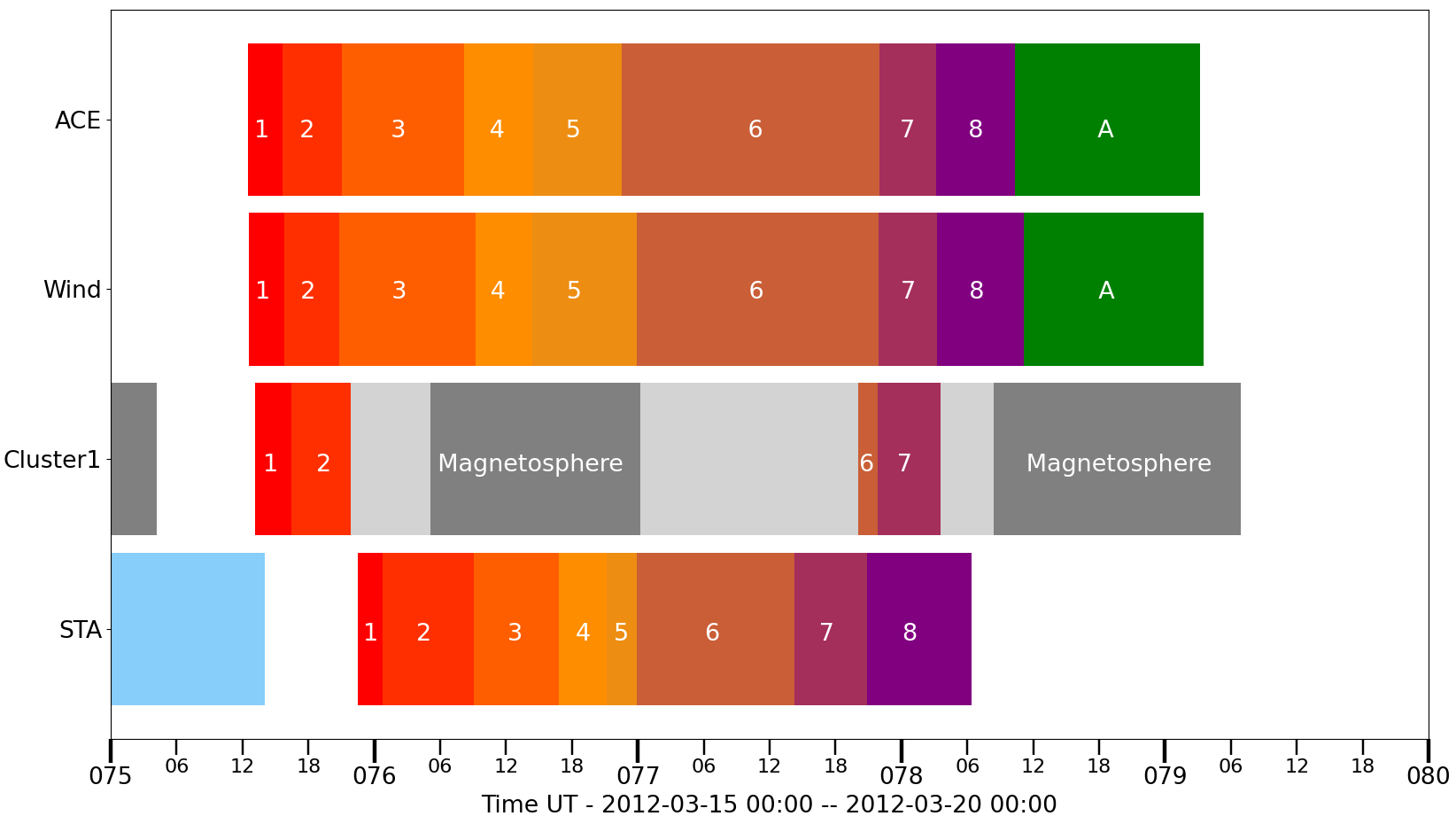}
	\caption{Timeline of the different identified crossed regions (see Section \ref{sec:in_situ} for the definition of the regions) for each spacecraft. Each colour represents each of them  (see text for more details). Note that Cluster did not observe the complete sequence of regions due to its entry into the magnetosphere (grey areas)} 
	\label{fig:timeline}
\end{figure}

A timeline of the identified crossed regions is shown in Figure \ref{fig:timeline}. Table \ref{tab:timeline} describes the different regions as well as lists the time interval when they were observed, the duration of their passage, and the comparison of the time span with respect to Wind observations. It should be noted that, as described in Section \ref{sec:in_situ}, those regions do not correspond one-to-one to the traditional identification of the different parts of an in situ measured CME (see Section \ref{sec:introduction}), but they are purely based on the SW speed tendency.

As mentioned above, the bulk SW speed profile shows extremely similar behaviours, although there is an almost-systematic offset of $\unsim200$ km/s less in the case of STEREO-A, most likely produced by the interaction of the CME with the preceding SIR \citep[see e.g.][for a similar case]{Heinemann2019}. The different regions have been defined based on the discontinuities and changes of tendency of this plasma property. A brief summary of the particular features observed in each region is also described at the bottom of Table \ref{tab:timeline}.

% \usepackage{booktabs}

% \usepackage{multirow}
\begin{table*}[]
\centering
\begin{tabular}{cccccc}
\hline
s/c & Region & \begin{tabular}[c]{@{}c@{}}Start (March 2012)\\ (day HH:MM UT)\end{tabular} 
 & \begin{tabular}[c]{@{}c@{}}End (March 2012)\\ (day HH:MM UT)\end{tabular} & \begin{tabular}[c]{@{}c@{}}Duration\\ (minutes)\end{tabular} & \begin{tabular}[c]{@{}c@{}}$\Delta t$ wrt\\ Wind (minutes)\end{tabular}\\ \hline
\multirow{9}{*}{Wind}     & \#1  & 15 12:34 & 15 15:48 & 194  & - \\
                          & \#2  & 15 15:48 & 15 20:46 & 298  & - \\
                          & \#3 & 15 20:46 & 16 09:14 & 748  & -\\
                          & \#4 & 16 09:14 & 16 14:21 & 308  & -\\
                          & \#5 & 16 14:21 & 16 23:52 & 570  & -\\
                          & \#6 & 16 23:52 & 17 21:57 & 1325 & -\\
                          & \#7 & 17 21:57 & 18 03:13 & 316  & -\\ 
                          & \#8 & 18 03:13 & 18 11:08 & 475  & -\\
                          & A & 18 11:08 & 19 03:30 & 982  & -\\
                          \hline
\multirow{9}{*}{ACE}      & \#1  & 15 12:32 & 15 15:40 & 188  & -6  \\
                          & \#2  & 15 15:40 & 15 21:03 & 323  & +25  \\
                          & \#3 & 15 21:03 & 16 08:12 & 669  & -79 \\
                          & \#4 & 16 08:12 & 16 14:27 & 375  & +67 \\
                          & \#5 & 16 14:27 & 16 22:32 & 485  & -85  \\
                          & \#6 & 16 22:32 & 17 22:00 & 1408 & +83 \\
                          & \#7 & 17 22:00 & 18 03:11 & 311  & -5  \\ 
                          & \#8 & 18 03:11 & 18 10:19 & 428  & -47  \\ 
                          & A & 18 10:19 & 19 03:09 & 1010  & +28  \\ 
                          \hline
\multirow{9}{*}{Cluster 1}& Magnetosphere & <15 00:00 & 15 04:12 & -  & -  \\
                          & \#1  & 15 13:08 & 15 16:27 & 199  & +5  \\
                          & \#2  & 15 16:27 & 15 21:53 & 326  & +28  \\
                          & Interaction & 15 21:53 & 16 05:09 & 436  & - \\
                          & Magnetosphere & 16 05:09 & 17 00:15 & 1146  & - \\
                          & Interaction & 17 00:15 & 17 20:04 & 1189  & -  \\
                          & \#6 & 17 20:04 & 17 21:50 & 106* & -* \\
                          & \#7 & 17 21:50 & 18 03:32 & 342  & 26  \\ 
                          & Interaction & 18 03:32 & 18 08:24 & 291  & -  \\ 
                          & Magnetosphere & 18 08:24 & 19 06:54 & 1351  & -  \\ 
                          % & Magnetosphere & 20 15:33 & 21 11:12 & 2004  & -  \\
                           \hline
                          
\multirow{8}{*}{STA} & \#1  & 15 22:29 & 16 00:44 & 135  & -59  \\
                          & \#2  & 16 00:44 & 16 09:04 & 314  & -16   \\
                          & \#3 & 16 09:04 & 16 16:48 & 463  & -350 \\
                          & \#4 & 16 16:48 & 16 21:11 & 263  & +20  \\
                          & \#5 & 16 21:11 & 16 23:56 & 165  & -405 \\
                          & \#6 & 16 23:56 & 17 14:14 & 858 & -467 \\
                          & \#7 & 17 14:14 & 17 20:54 & 400  & +84 \\ 
                          & \#8 & 17 20:54 & 18 06:22 & 568  & -414 \\ %\hline

\end{tabular}
\begin{tabular}{cp{0.9\textwidth}}
\hline
Region & Description\\ \hline
\#1  & It starts with the shock and forms part of the sheath region.\\
\#2  & It is defined between a sudden enhancement of the bulk speed and the sudden depletion of this physical property. It is partially covering the sheath, and from the near-Earth observations, part of the magnetic obstacle of the structure too.\\
\#3 & Expanding region. It has a clear FR structure at Earth, but not at STEREO-A. For Earth, it can be considered as a magnetic cloud \citep{teresa}, while for STEREO-A is considered part of the sheath \citep{Lan_Jian_1}.\\
\#4 & Increasing speed from $\unsim625$ to $\unsim700$ km/s at Earth, and from $\unsim425$ to $\unsim500$ km/s at STEREO-A. \\
\#5 & The speed profile presents a "u" shape or parabola behaviour. \\
\#6 & Speed remains more or less constant with a very fluctuating behaviour. \\
\#7 & It shows a gradual decrease of the speed with smooth behaviour for the three s/c, indicating an expansion. The magnetic field strength matches for all the observers.\\
\#8 & Expanding region. \\
A & Expanding region ending with a bulk speed plateau that ends up in a SIR. (Observed only near Earth)\\
Interaction & It is a mixture of the magnetopause and the transit of the in situ CME. \\\hline
\end{tabular}
% \vspace{0}
\caption{Start, end, and duration of the different regions of the in situ measured CME based on changes in the tendency of the SW speed for the different spacecraft. Last column compares the duration of the different regions at each spacecraft with respect to Wind.}
\label{tab:timeline}
\begin{flushleft}
    {*Comparison to Wind is not possible due to the previous interaction with the magnetosphere.} 
\end{flushleft}%\raggedright \par}
\end{table*}

\begin{figure*}  
   \centering
	\includegraphics[width=1\textwidth]{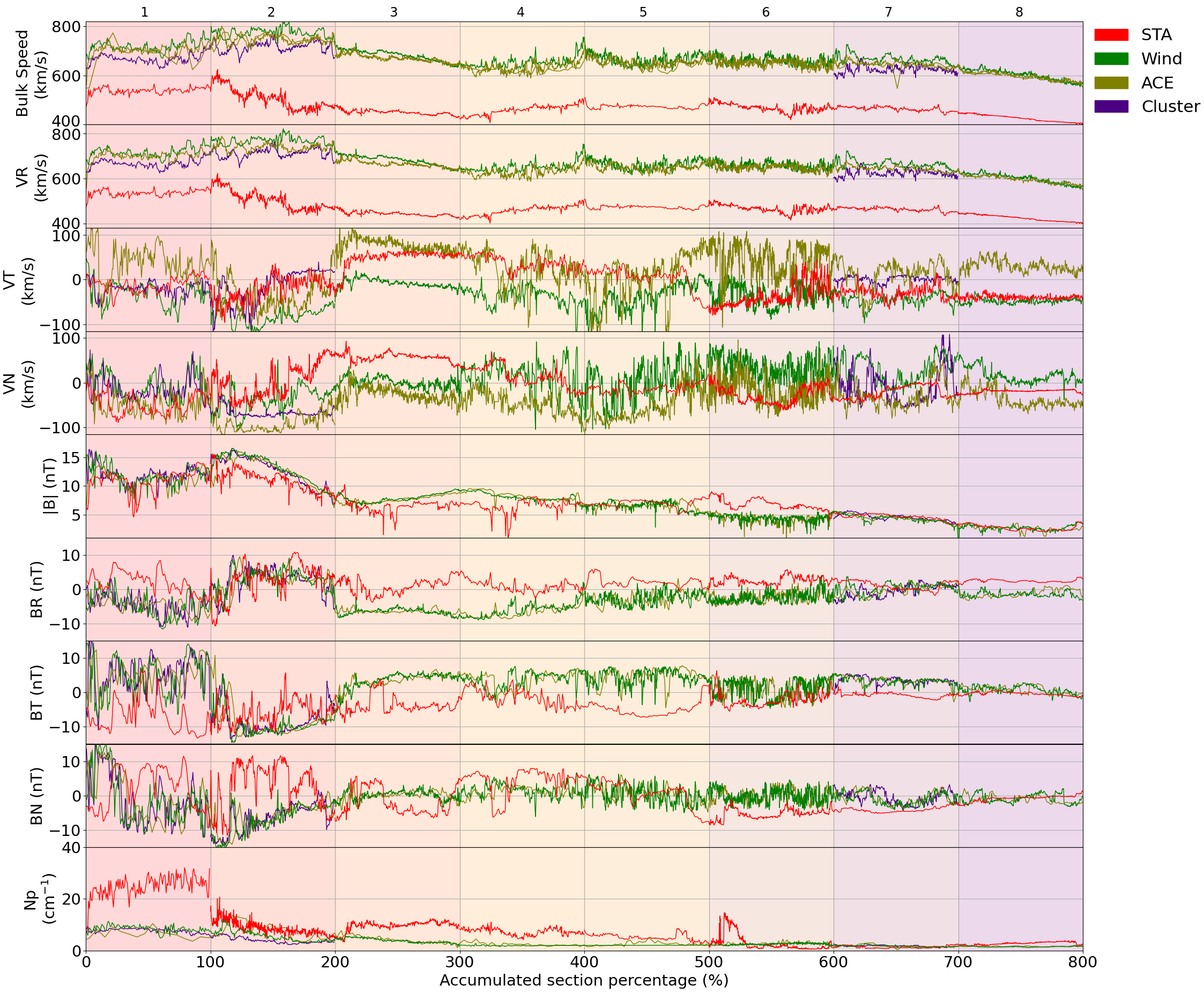}
	\caption{Comparison of the plasma and magnetic field properties through the different regions. From top to bottom: plasma velocity (magnitude and RTN components), magnetic field (magnitude and RTN components), and proton density. X-axis corresponds to the cumulative percentage of data points of the different regions. Each shade colour corresponds to the different regions utilised in previous figures. All data are resampled to a 1-minute cadence. See text for more details.}
	\label{fig:superposed_epoch}
\end{figure*}

Considering these different regions, an analysis comparing the plasma properties of the passage of the CME through each spacecraft has been performed using normalisation of the time series to initial and end times of observed structures for each region. Figure \ref{fig:superposed_epoch} shows from top to bottom: the bulk SW speed, the components of the velocity in RTN coordinates ($V_R$, $V_T$, $V_N$), the magnetic field magnitude, the components of the magnetic field vector in RTN ($B_R$, $B_T$, $B_N$), and the proton SW density. The red traces indicate STEREO-A observations; dark green, Wind observations; olive green, ACE; and purple, Cluster-4. The SW speed and magnetic field strength exhibit an extraordinary similarity at the different locations and throughout all the regions despite the lower values (by about $\unsim200$ km/s) measured by STEREO-A, and some small differences in the field and velocity components. We outline the following details when comparing the measurements of the spacecraft:

\begin{itemize}
    \item The bulk SW speed obviously displays similar behaviour during all regions, as this parameter was used to select the boundaries. However, a systematic offset of $\unsim200$ km/s between STEREO-A and Earth can be observed.
    \item Most of the speed corresponds to the radial component, as it is the predominant direction of the SW. In this case, the velocity in the T-N plane would be representative of the lateral expansion of the structure \citep{expansion_cmes}. Regions \#2, \#3, \#7 and \#8 show similar behaviour of the velocity components. 
    \item The magnetic field magnitude is very similar during the entire transit, especially during regions \#1, \#2, \#7 and \#8.
    \item The radial and tangential components of the magnetic field vector display both similar values and tendency during region \#2, and stable values with similar behaviour during regions \#7 and \#8.
    \item Regions \#4, \#5, \#6 present fluctuating velocity components and less correlated magnetic field strength and components. These differences, together with the similarities of the surrounding areas, suggest that these regions (\#4, \#5, \#6) are under different plasma conditions. Also, it supports the idea that at both locations (i.e. STEREO-A and Earth), the main structure is being crossed twice (in, out, in. See Section \ref{sec:scenario} for more details).
    \item In all regions, the proton density shows a clear difference between STEREO-A and Wind, being almost systematically $\unsim3$ times greater at STEREO-A, most likely due to the interaction with the SIR, which also slows down the CME propagation in the Sun--STEREO-A direction.
\end{itemize}

As shown in Figure \ref{fig:superposed_epoch}, there are certain parts during the transit of the CME through the spacecraft with such similarity that evinces the presence of the same structure crossing STEREO-A and near-Earth observatories despite their notable separation.

\begin{figure}[htbp]  
   \centering
	\includegraphics[width=\columnwidth]
    {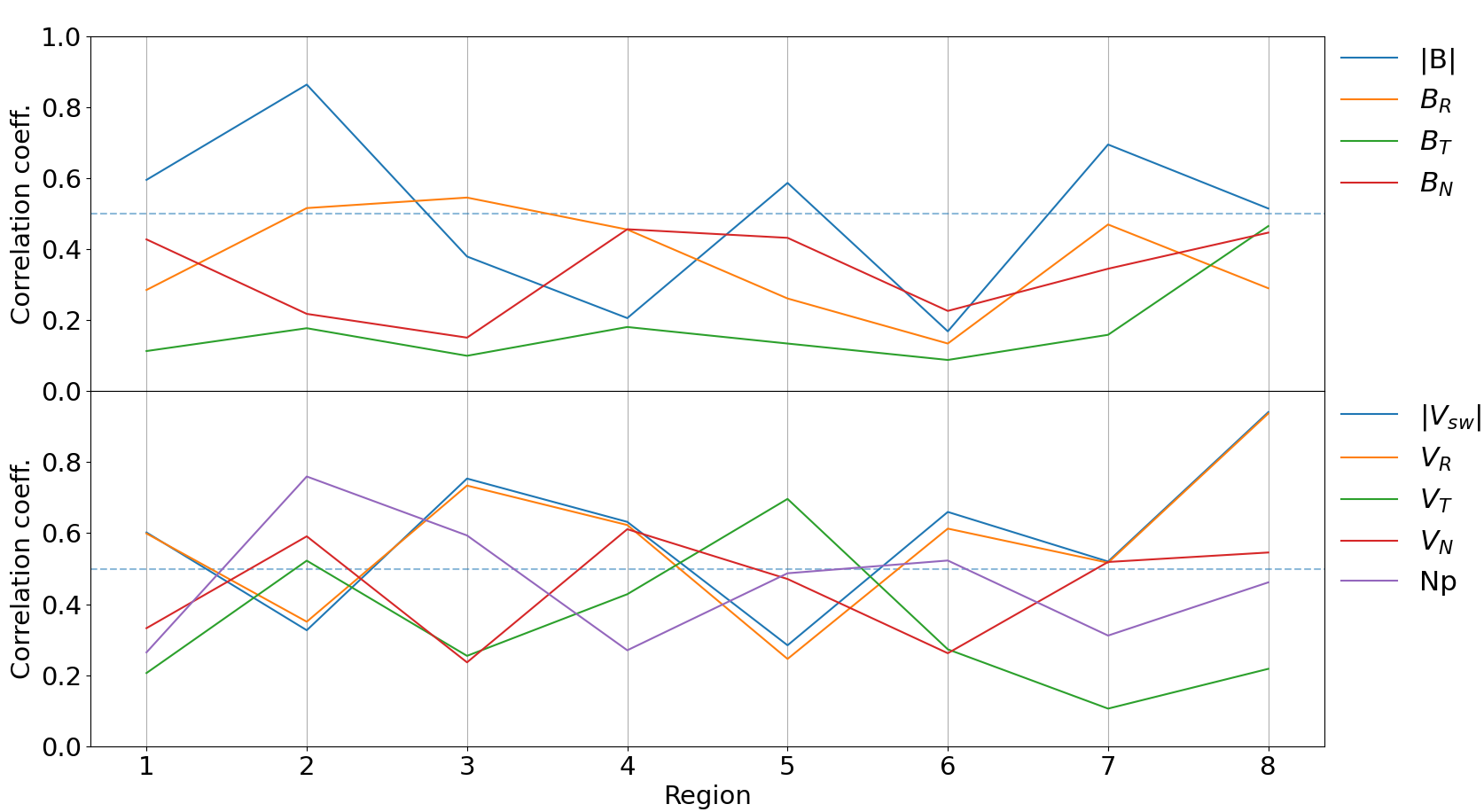}
	\caption{Greatest Pearson correlation coefficient values found between STEREO-A and near-Earth spacecraft observations for each defined sector of the parameters shown in Figure \ref{fig:superposed_epoch}.}
	\label{fig:correlations}
\end{figure}

% In order to quantify the similarity between the SW speed, velocity components, density, magnetic field magnitude and its components at STEREO-A and near-Earth in each region, we have performed a linear regression among the parameters observed by
In order to quantify the similarity between the SW speed, velocity components, magnetic field magnitude, magnetic field components, and SW density at STEREO-A and near-Earth in each region of the CME passage, Figure \ref{fig:correlations} shows the Pearson correlation coefficient between them computed for each region. The calculation has been performed correlating the observations of each in situ parameter from STEREO-A and near-Earth spacecraft. Then, the largest correlation out of the three combinations (STEREO-A and Wind, STEREO-A and ACE, or STEREO-A and Cluster) has been selected. Despite the long separation between STEREO-A and near-Earth spacecraft, local particularities, and the evolutionary processes undergone by the structure along the different directions (apart from the instrumental differences and cadences), the speed and magnetic field strength are correlated with a Pearson correlation coefficient greater than 0.5 for most of the crossed regions. In particular, the magnetic field displays a better correlation during the front (regions \#1 and \#2) and rear portions of the CME (regions \#7 and \#8), while the SW speed does for regions \#1, \#3, \#4, \#7 and \#8. Although the main property used for selecting the boundaries of each region was the bulk speed, the correlation is higher for the magnetic field. The above results suggest that the structure is magnetically coherent despite the local plasma properties. 

Apart from the difference in time between ACE and Wind observations of the different features of the bulk speed profile (that can be more than one hour apart. See Table \ref{tab:timeline}), there is also a difference in the velocity components despite their relatively small separation (see Figure \ref{fig:superposed_epoch}). This may indicate that the expansion of the structure (regions \#3 and \#8) is occurring differently at both positions, probably due differences in the proximity of both spacecraft to the central part of the expanding local plasma %, as well as the quadrants of the trajectory from a hypothetical local reference system 
\citep[i.e. the impact parameter of both spacecraft; see e.g.][for details]{Dasso2007, expansion_frs}. % When the impact parameter (p/R) is small... 
The quasi-stable values and the increment of the tangential (T) and normal (N) components with respect to the radial (R) could be also indicative of the existence of a transit through the leg of the structure \citep[as derived from][]{expansion_frs}. %Owens?
% \textcolor{red}{MAD -- do you mean an increase of values of non-radial flows indicate leg crossing? }

Besides, considering the interaction of the structure with the locally encountered upstream ambient SW, the piled-up plasma in the sheath (region \#1) presents a similar behaviour and an unexpectedly high correlation, considering the completely different conditions of the SW that the structure suffered while propagating to the observatories.

\subsubsection{Suprathermal Electron Pitch-angle Distributions}\label{sec:pad}

\begin{figure}[htbp]  
   \centering
   \includegraphics[width=\columnwidth]{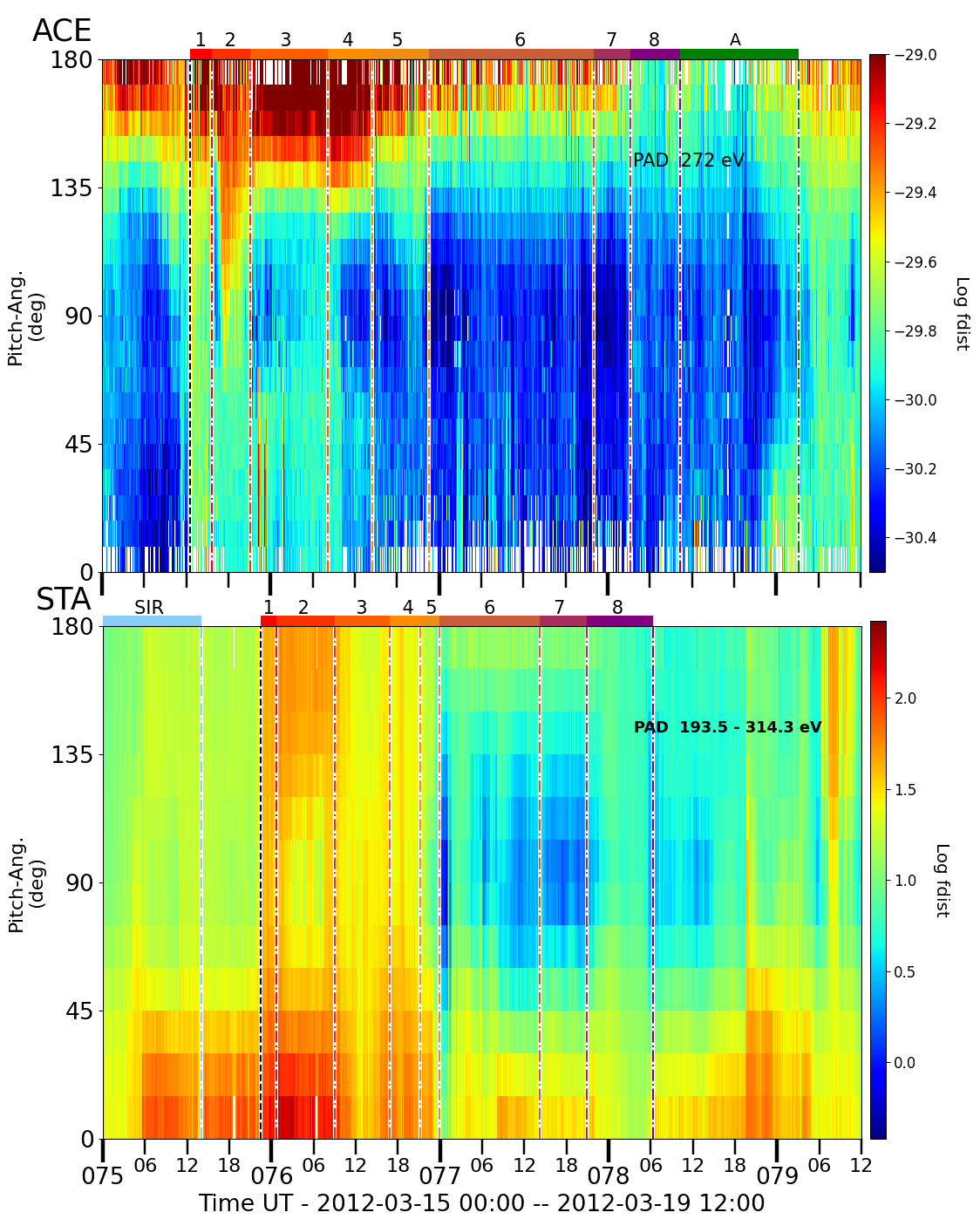}
    \caption{\textbf{Top:} Suprathermal electron PADs (time series of the distribution function, Fdist, with respect to the magnetic field angle) for the 272 eV energy channel measured by ACE/SWEPAM from 2012-03-14 to 2012-03-19 at 12:00 UT. \textbf{Bottom:} In situ PADs of the suprathermal electrons from 193.5 to 314.3 eV measured by STEREO-A/SWEA for the same period. Dashed lines and top colour bars represent the different defined regions. See text for more details. %\textcolor{red}{Probably replace ACE by Wind. ADD NUMBERS OF REGIONS. Lynn -- At the very least, add it as another panel here... Lan -- If you merge this with Fig 5 and 6, you can add Wind suprathermal electron PAD too. Since this figure is mentioned when describing Fig 5, you may want to move the figure forward if you don't merge it.}
    }
	\label{fig:pads}
\end{figure}

Suprathermal electrons are a good tracer of the interplanetary magnetic field topology and properties, especially when comparing their pitch-angle with respect to the direction of the magnetic field. Figure \ref{fig:pads} top (bottom) panel shows the 272 eV (193.5--314.3 eV) PADs obtained from ACE/SWEPAM (STEREO-A/SWEA) measurements from 00:00 UT on March 15 to 12:00 UT on March 19. Different colour blocks on top of each panel and dashed lines correspond to the previously defined regions (see Section \ref{sec:cross_section_comparison}). The sporadic peaks (and floor) are intentionally saturated, as their high (low) intensity smears the visual contrast of the rest of the period.

Focusing on the top panel, ACE observed a simple \textit{strahl} coming from the anti-parallel direction before the arrival of the CME. The width of the \textit{strahl} becomes broader while approaching the CME shock. Within region \#1, the PAD is nearly isotropic, which is a typical signature of the sheaths \citep[e.g.][]{carcaboso2020}. Region \#2 displays periods of BDE especially at the beginning and the \textit{strahl} electron width coming from 180\degree{} becomes larger until the end of the period. This behaviour suggests there is a still linked-to-the-Sun tube (i.e., a FR) of particles which is undergoing local disturbances that broaden the \textit{strahl} predominantly coming from 180\degree{} \citep[see e.g.][for potential sources of scattering]{energy_focusing2}. The reason underneath this observation may result from a local feature within region \#2 because the smoothness of the magnetic field, in this overall expanding region, will cause the electrons to focus more easily than in the ambient SW due to adiabatic expansion \citep{strahl_var}.%FCM: One of the possibilities might be the interaction with a part of the Earth's magnetosphere. https://agupubs.onlinelibrary.wiley.com/doi/full/10.1029/2011JA017269 : As a result, the relative decrease in field magnitude along an ICME field line between the coronal source region and 1 AU may be greater in ICME than in non-ICME solar wind, thus driving stronger magnetic focusing within ICMEs.

The top panel of Figure \ref{fig:pads} shows that region \#3 was characterised by clear BDEs, with the electrons coming from 0\degree{} presenting larger width but less intensity than those coming from 180\degree{}. The intensities of the electrons coming from 0\degree{} decrease in region \#4, and the width of those coming from 180\degree{} increases. Region \#5 has a similar behaviour to region \#2 in the predominant \textit{strahl}. Region \#6 shows sporadic periods of BDE and isotropy, and predominant simple \textit{strahl}, which suggests that there is a mix of open, closed, and completely detached magnetic field lines. Overall, region \#7 shows a simple \textit{strahl} narrower than in previous regions. Region \#8 shows a weaker intensity of those electrons coming from 180\degree{}. Also, it displays some intermittent periods of BDE and isotropy. Finally, region A shows simple \textit{strahl} with some isotropic periods. At the end, the counterstreaming electrons are becoming clearer due to the proximity of the upcoming SIR.

The whole transit of ACE through the structure presents periods of more or less clear BDEs (sometimes longer, sometimes patchy behaviour), and systematically, a \textit{strahl} population with pitch angles $\unsim0\degree{}$ less intense than the population with pitch angles $\unsim180\degree{}$. This could be related to having a flank crossing close to one leg of the structure, which apparently owns a preferential negative (inward) direction \citep[see examples in][]{strahl_leg}. In addition, it agrees with the remotely-observed handedness (left) previously mentioned in Section \ref{sec:cme} and reported by \citet{erika2018}, having the more intense \textit{strahl} population in the opposite direction than the dominant magnetic field. As discussed below, this is consistent with the spacecraft crossing the east leg of the structure. On the other hand, the absence of BDEs may be an indicator that only part of the CME remained magnetically closed, whereas the other parts were eroded (possibly due to interaction with the fast SW) or originally were open magnetic field lines \citep[e.g.][]{winslow_multipoint, carcaboso2020}. 

The bottom panel of Figure \ref{fig:pads} shows STEREO-A observations. The blue-shaded period marked in Figure \ref{fig:sta_insitu} is also in the bottom panel of Figure \ref{fig:pads}, and corresponds to the passage of a SIR that started to cross STEREO-A on March 14, as identified by \cite{sir1}. The difference between the \textit{strahl} peak and the background is not as pronounced as it is during the passage of the high-speed stream right before the arrival of the CME. The \textit{strahl} population of the suprathermal electrons follows the polarity of the nominal Parker spiral (positive). Within the CME passage at STEREO-A, region \#1 displays isotropic electron PADs, similarly to ACE; region \#2 shows clear BDEs, with some periods of isotropy; the boundary between region \#2 and \#3 is completely isotropic, and region \#3 shows a smeared \textit{strahl} population, reaching widths greater than 90\degree{} and reaching the complete isotropy by the end, indicating the presence of magnetic in situ reconnections \citep[e.g.][]{iso_3}; region \#4 shows a similar behaviour, but with clearer isotropy; region \#5 gradually shows greater BDEs, and regions \#6 and \#7 shows the clearest BDE period, with higher intensities and \textit{strahl} widths at 0\degree{} than 180\degree{}; finally, region \#8 also shows BDE but with less gradient between peaks and background.  %The change of pitch-angle at 21:30 UT on March 14 indicates the crossing of the heliospheric current sheet before the arrival of the SIR. 

The overall tendency of the electron PADs at STEREO-A during the transit of the CME is the clear presence of BDEs with a predominant peak intensity parallel to the magnetic field. Assuming that the magnetic field is predominantly axial, and STEREO-A crossed closer to the west leg of the CME, this behaviour supports the idea that the handedness derived from remote sensing observations (left-handedness) previously reported by \cite{erika2018} also matches the one observed in situ. % It is more likely to be related to other events than this one...... Approximately 8 months after the eruption (2012-11-24 20:00), a considerable enhancement of the magnetic field (from 380 pT to 593 pT, being the maximum value on 2012-12-22 13:00), likely to be associated with a shock, is captured by the magnetometer onboard Voyager 1 \citep{voyager}. At the beginning of the shock, the spacecraft was located at 125.75 au, 34.5 degrees in latitude and 130 degrees in longitude (Stonyhurst). Voyager 2 though does not see any significant signature of a transitory structure.
In this case, the peak intensity in the BDEs could be also indicative of the proximity to the different legs of the structure \citep{strahl_leg}. While STEREO-A observes a more intense flux coming from the parallel direction (pitch-angle $\unsim0\degree{}$), near-Earth measurements show a predominant higher peak with pitch angles at $\unsim180\degree{}$ (anti-parallel to the magnetic field).

% Apart from that, the difference between the two peaks is more pronounced in the case of Earth's observations.

Considering the longitudinal separation between STEREO-A and Earth, the suprathermal electrons would take a longer time to travel from one location to the other than the actual passage time of each defined CME region over each spacecraft. As a reference, a completely field-aligned electron of 300 eV would take approximately 13.5 hours to travel 1.62 au, which corresponds to the absolute distance between STEREO-A and Earth (actual magnetic field path-length would be longer in a FR structure, and the particle would have an inherent gyroradius). Despite this fact, there are remarkable similarities in their PADs at both locations, suggesting that the transport conditions that those electrons may have experienced are similar regardless of having travelled completely different paths. The different regions (purely defined by the changes in the SW speed tendencies) are strongly linked to the behaviour of the PADs.

\section{Possible Scenario}\label{sec:scenario}

\begin{figure}[htbp]  
   \centering
	\includegraphics[width=\columnwidth]{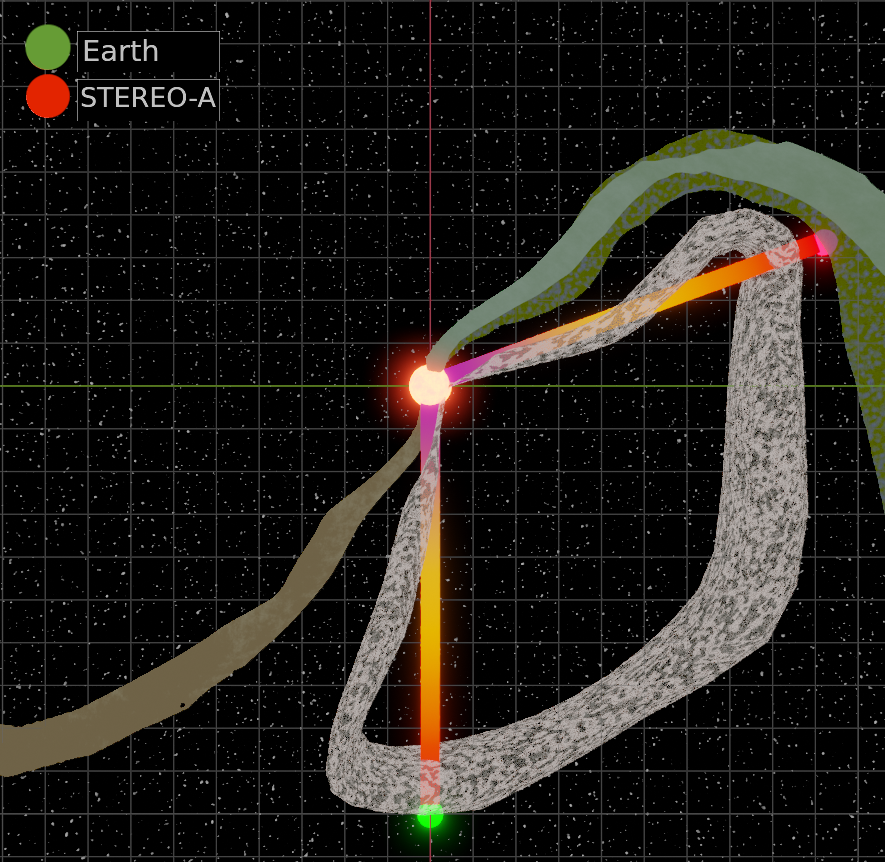}
    \includegraphics[width=\columnwidth]{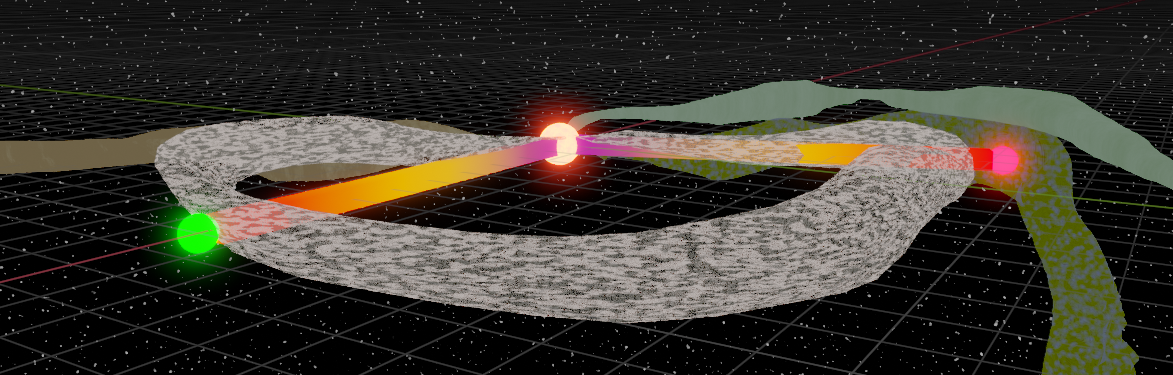}
	\caption{Proposed scenario based on the observations. Spacecraft and structures are located at the moment of the encounter with Wind. At that moment, STEREO-A was found in the high-speed stream. The green and brown traces represent the high-speed streams generated by CH2/CH3 and CH1, respectively. Colour bars represent the radial cross through the conditions of the SW following the same format as \text{Figures \ref{fig:timeline}} and \ref{fig:superposed_epoch}.} 
	\label{fig:scenario}
\end{figure}

Figure \ref{fig:scenario} sketches the envisaged scenario of the CME structure and the SW conditions in the interplanetary medium during the event. This scenario is based on the measurements and the performed analysis and represents the most plausible reconstruction according to our assessment. However, other alternative interpretations could exist. The top panel represents a view from the north of the ecliptic and the bottom panel a view from the nose of the structure directed between Earth (green sphere) and STEREO-A (red sphere). Earth would cross the east flank of the CME \text{$\unsim10$ hours} before STEREO-A, which would cross the west flank. At the time Earth crosses the shock and sheath of the structure, STEREO-A is crossing a high-speed stream (indicated by the greenish tube in Figure \ref{fig:scenario}) likely originated from CH3 (or the combination of CH2 and CH3).

According to the GCS reconstructions, the CME originated $\unsim55\degree{}$ in longitude between both spacecraft whereas the half angle of the structure was $\unsim50\degree{}$. The structure was also very inclined ($\unsim45\degree{}$ with respect to the ecliptic plane), so in order to cover a large longitudinal span over the ecliptic plane later at \text{1 au}, the structure had to widely expand in that direction, rotate and/or bend to cover that space. Shock orientations (at Earth pointing southwest, and at STEREO-A pointing west) indicate that the deformation/rotation was very likely.

There are also indications that the portion of the CME directed toward STEREO-A would have started to interact with the high-speed streams produced by CH2 and CH3 at closer radial distances% \textcolor{red}{Lan -- not necessarily SIRs (which requires slow-to-fast wind interaction), but CME should have been affected by the intermediate-speed solar wind from CH2 and CH3, because these two CHs are not very large. }
, such as the presence of higher densities at STEREO-A (3 times greater than at Earth), the speed was $\unsim200$ km/s lower, as well as the duration of the passage (shorter in the case of STEREO-A), suggesting that the structure was likely more compressed on average in that portion of the CME.

% manuela.temmer: that is very instructive and finally helped me to understand all the results before. maybe to introduce that at an earlier stage as hypothesis and then guide a reader towards that scenario by observational support. Mar 30, 2023 12:19 PM

The similarities of different plasma properties at the two locations (bulk SW speed, behaviour of the velocity components, magnetic field strength, PADs...) suggest that at least the three first regions (\#1, \#2, \#3) and the two last regions (\#7, \#8) correspond to the same structure, which matches with the CME originated on March 13 at 17:12 UT, while regions \#4 and \#5 display mixed signatures of ambient SW and magnetic obstacle, with evidence of local reconnection as indicated by, among others, the observed suprathermal electrons PADs. Additionally, region \#6 in the case of Earth's observations does not show clear signatures of being part of the core of the structure, while for STEREO-A, there is a clear inclined and smooth magnetic field, with low temperatures, low plasma $\beta$ and BDEs, for which it could be identified as a MC.

This scenario differs from what \cite{contradicting_paper} found for this event based only on near-Earth in situ observations (see Figure 5 in that article). That study proposes that from region \#4 to \#8, the SW presents high-speed stream signatures. From our observations, we did not find clear evidence for the source of this fast SW, but the one produced by the CME itself.

% \textcolor{red}{CME at STA is more compressed than at Earth}
% \textcolor{red}{The structure must deform to cover the ecliptic plane. Shock normal for STA in the ecliptic plane, for Wind, pointing south}

% Shock normal STA: (-0.079, -0.997, -0.002)
% Shock normal Wind: (-0.771, -0.538, -0.340)

\section{Summary}\label{sec:discussion}

% \textcolor{red}{AV -- The Discussion section lacks coherence. And it would be good to also have a short conclusions section, on what is the main outcome and interpretation of the study,}

A CME associated with a M7.9 flare was released on March 13, 2012, with a speed of $\unsim2000$ km/s, and an inclination of approximately 45\degree{}. The CME started propagating outward from the Sun in a direction 55\degree{} east from Earth, i.e. between Earth and STEREO-A which were separated by $\unsim110\degree{}$ apart from each other in longitude. The AR was surrounded by three CHs. At least one of them interacted with the CME while it propagated away from the Sun, hampering its propagation towards STEREO-A.

The GCS reconstruction and the derived kinematics seem to be consistent with the arrival time of the CME at both locations and the in situ measurements at $\unsim1$ au. There is no evidence of the presence of other CMEs of similar magnitude 24 hours prior to and after the CME, observed by LASCO and SECCHI coronagraphs, which erupted on March 13, 2012 at $\unsim17:30$ UT. This can be assured thanks to the orbital position of the two STEREOs and Earth, which allowed a 360\degree{} visual coverage of the Sun.

The DBM model suggests that although it is the same structure, there is a $\unsim46{\%}$ more drag towards STEREO-A than towards Earth, most likely because of two complementing effects: a previous CME towards Earth swept the SW in that direction, and the presence of the high-speed streams produced by CH2 and CH3. % \textcolor{red}{AV -- This needs more elaboration. In the Results section and here.}

The computation of the shock parameters presented indicates that the shock normal vector at Wind location was pointing southwest and, at STEREO-A location, it was contained in the ecliptic plane pointing west. %\textcolor{red}{This is not a paragraph or even a complete sentence.  Regardless of that, the orientation of the shock normal unit vector should not be terribly surprising if the ICME is not a perfect sphere (which it almost certainly is not) with origin propagating in plane created by the Sun-Wind-STEREO set of points.}

Despite the large longitudinal separation and initial inclination of the structure, the bulk SW speed and magnetic field magnitude displayed almost identical profiles on near-Earth observatories and STEREO-A, suggesting that the observed CME is the same structure at both spatial locations. STEREO-A transited the west flank, while Earth did it through the east one. The in situ observations show opposite field polarity on STEREO-A compared to Wind/ACE (negative for Earth’s and positive for STEREO-A’s), and the suprathermal electron PAD during the BDE periods supports the in situ left-handedness, matching the one reported by \cite{erika2018} using remote sensing observations. Also, the higher intensity of the \textit{strahl} population at 0\degree{} on STEREO-A, and at 180\degree{} on Earth’s observation, during the BDE periods could indicate the proximity to the magnetic foot point of the CME. % \textcolor{red}{Lan -- I think it is because the polarities of the underlining IMF sectors are different} %This paper evidences an example of the mixture of ambient SW together with the MO of the CME. The reconnection that may happen to produce that mixture is not enough to smear completely the BDEs. %The effects can extent up to

As it can be derived from the timeline and the $\Delta{t}$ column in Table \ref{tab:timeline}, some parts of the structure seem to have transits of different duration depending on the observer (even with the very-close-to-each-other Wind and ACE). This has significant implications for our understanding of the CMEs, as most of the studies are based on just single point observations.

Despite not having clear smooth magnetic field rotation in the STEREO-A region \#2, there are pronounced BDEs, similar to those observed at ACE, but with the highest peak in the opposite direction (suggesting that both spacecraft intercepted the flanks of the CME close to the respective legs, as mentioned above). Assuming that the regions are part of the same structure and that no remarkable evolutionary processes have happened (which may have resulted in a different imprint in the PADs), the identification of what traditionally has been called FR extends further than the actual helical magnetic field, which would correspond to the magnetic obstacle or region \#3.

A geomagnetic storm with Kp > 5 was produced while Earth was crossing a strong negative $B_z$ component. The geomagnetic effects (Kp $\unsim4$) lasted until the end of the passage of the structure by Earth. However, previous catalogues reported a shorter in situ CME passage.

\section{Conclusion and Discussion}\label{sec:conclusion}

The main conclusions of this work can be stated as follows:
\begin{enumerate}
    \item There is clear evidence of having the same CME observed by 110\degree{} separation although its initial inclination was $\unsim45\degree{}$. Based on these observational facts, we propose a scenario (Section \ref{sec:scenario}) in which an originally highly inclined wide CME deformed and/or rotated in such a way that it was observed near its legs by spacecraft in the ecliptic plane separated by >100\degree{} in longitude. 
    \item The structure interacts with at least one of the high-speed streams produced by the west CHs (CH2, CH3) surrounding the AR, which probably caused the deformation and deceleration of the structure in the STEREO-A direction. 
    \item The piled-up plasma observed in the sheath (region \#1) has an unexpectedly extraordinary correlation in the magnetic field strength at the two locations separated by 110\degree{} in longitude. This plasma is supposedly a mixture of the CME and the local ambient SW compressed downstream the shock. Despite the ambient conditions at both locations were different, the similarities in the magnetic field strength require a deeper analysis.
    \item The in situ observations reveal that at both locations each spacecraft crossed the respective flanks of the CME, then ambient SW, and finally the legs' proximity of the CME. In the case of STEREO-A, the cross through the leg is more evident than for near-Earth spacecraft (See Figure \ref{fig:scenario}, which sketches the proposed scenario).%there is a cross through the CME, a central part which shows signatures of ambient SW, and a cross through both legs of the structure. 
    \item We found no clear evidence for a potential source for a high-speed stream crossing Earth during the period from March 16 to March 18 as suggested by \cite{contradicting_paper}. %That period would (partially) still correspond to the transit of a CME as explained in Section \ref{sec:scenario}.
    
\end{enumerate}

% Some parts of the CME are compressed, while others are expanded

% \textcolor{red}{Lo que vengo a decir es que hay diferencias significativas entre por ejemplo ACE and Wind, o en el evento de Noé donde están relativamente cerca y ven cosas distintas, pero aun así en este caso se parece mucho a STEREO-A...}

% Then we characterise the coronal holes.

% We use multiple methods to infer what is going on

% We analyse the in situ plasma properties to support the scenario

% \textcolor{red}{MAD -- I would put this part in discussion and only make a short sentence referring to it here in the conclusions} 

The similarities of the structure regardless of the large angular separation are evidenced, in contrast to previous studies, such as the case-study event analysed by \cite{lugaz_55}, where less clear correlations are found despite the spacecraft locations observing the CME were closer to each other. This shows that the factors that determine each particular  observation (i.e. how the spacecraft intercepts the CME, the morphology and evolutionary processes of the CME, SW conditions, etc.) play the main role in the interpretation of the observations and derived scientific results%\textcolor{red}{AV -- This sounds vquite unspecific}
. On the other side, having relatively close proximity does not assure similar measurements. They may significantly differ more than one hour in the definition of each region based on the bulk speed (e.g. see column $\Delta{t}$ for near-Earth spacecraft on Table \ref{tab:timeline}). These differences play a very relevant role in the scientific explanation, and thus in our understanding of the large-scale structures, such as the CMEs are. Even with the extensive catalogues available, the use of single spacecraft data during the transit of the CMEs implies that the derived interpretations may suffer large inconsistencies. Nevertheless, the use of the available multipoint measurements helps, and it is even required, to support the different encountered scenarios.

This case study opens up questions about our current view of the CME longitudinal spread and its evolution in the interplanetary space, pointing out the need to find more examples of widespread CMEs and analyse them. Such analysis will help us understand the eruption, evolution and global topology of CMEs, as well as improve our modelling efforts.

\begin{acknowledgements}
          The authors acknowledge the open policy of the data for the different missions. We acknowledge use of NASA/GSFC's Space Physics Data Facility's CDAWeb service. F.~C. acknowledges the financial support by an appointment to the NASA Postdoctoral Program at NASA Goddard Space Flight Center, administered by Oak Ridge Associated Universities through a contract with NASA and the support of the Solar Orbiter mission. FC and LKJ thank the support of the STEREO mission. M.D. acknowledges the support by the Croatian Science Foundation under the project IP-2020-02-9893 (ICOHOSS). RGH acknowledges the financial support of the Spanish Ministerio de Ciencia Innovación y Universidades Project PID2019-104863RBI00/AEI/10.13039/501100011033 and by the European Union’s Horizon 2020 research and innovation program under grant agreement No. 101004159 (SERPENTINE). SGH acknowledges funding by the Austrian Science Fund (FWF): Erwin-Schr\"odinger fellowship J-4560. This work has been possible thanks to the scientific discussions within the LASSOS-Goddard group. F. C. thanks Benjamin Lynch, Angelos Vourlidas and Andreas Weiss for the discussions, and also the referee for their feedback. %This paper uses data from the Heliospheric Shock Database, generated and maintained at the University of Helsinki.
\end{acknowledgements}


\bibliographystyle{bibtex/aa}
\bibliography{bibtex/bibliography}

\begin{appendix}
\section{Missions and Instruments}\label{sec:instrumentation}
The instruments utilised in this study are listed below. On the one hand, the remote-sensing instrumentation listed by each mission was:

\begin{itemize}
    \item \textbf{Solar Dynamics Observatory} \citep[SDO,][]{sdo}: Atmospheric Imaging Assembly \citep[AIA,][]{sdo_aia}.
    \item \textbf{Solar and Heliospheric Observatory} \citep[SOHO,][]{soho}: Large Angle Spectroscopic Coronagraph \citep[LASCO,][]{soho_lasco}.
    \item \textbf{STEREO}: Heliospheric Imager \citep[HI,][]{stereo_hi}, Extreme UltraViolet Imager (EUVI), and the two coronagraphs (COR1, COR2), all being part of the suite Sun–Earth Connection Coronal and Heliospheric Investigation \citep[SECCHI,][]{stereo_cor1, stereo_secchi}. 
\end{itemize}

\noindent On the other hand, the in situ instrumentation was:

\begin{itemize}
    \item \textbf{ACE}: The Electron, Proton, and Alpha Monitor \citep[EPAM,][]{ace_epam}, the Solar Wind Electron Proton Alpha Monitor \citep[SWEPAM,][]{ace_swepam}, the Solar Wind Ion Composition Spectrometer \citep[SWICS,][]{ace_swics}, and the magnetometer of the Magnetic Fields Experiment \citep{ace_mag}.
    \item \textbf{Cluster}: The magnetometer \citep{cluster_mag} and the Cluster Ion Spectrometry experiment \citep{cluster_cis}.
    \item \textbf{STEREO-A}: The suite of instruments of the In situ Measurements of Particles And CME Transients \citep[IMPACT,][]{stereo_impact}, particularly the instruments Solar Wind Electron Analyzer \citep[SWEA,][]{stereo_swea} and the magnetometer \citep{stereo_mag}, and the Plasma and Suprathermal Ion Composition \citep[PLASTIC,][]{stereo_plastic}.
    \item \textbf{Wind}: The Solar Wind Experiment \citep[SWE,][]{wind_swe}, the Magnetic Field Investigation \citep[MFI,][]{wind_mfi}, and the 3-D Plasma and Energetic Particle Investigation \citep[3DP,][]{wind_3dp}.
\end{itemize}

\end{appendix}

\end{document}